\documentclass[twocolumn,twocolappendix]{aastex63}
\usepackage[utf8]{inputenc}

\usepackage{aas_macros,amsmath}
\usepackage{multirow,color}
\usepackage{amssymb}
\usepackage[T1]{fontenc}

\def\farcs{%
 \mbox{%
  \kern  0.13ex.%
  \kern -0.95ex\arcsec%
  \kern -0.1ex%
 }%
}%

\def\tcb{\textcolor{black}}

\newcommand{\oiii}{[O\,{\sc iii}]}
\newcommand{\oii}{[O\,{\sc ii}]}
\newcommand{\cii}{[C\,{\sc ii}]}
\newcommand{\nii}{[N\,{\sc ii}]}

\newcommand{\hst}{{\it HST}}
\newcommand{\jwst}{{\it JWST}}

\submitjournal{ApJ}
\shorttitle{
Resolving Chemical Enrichment at $z=4-6$
}
\shortauthors{Fujimoto et al.}

\begin{document}

\title{
The ALPINE-CRISTAL-JWST Survey: NIRSpec IFU Data Processing and \\
Spatially-resolved Views of Chemical Enrichment in Normal Galaxies at \boldmath $z=4-6$
}

\correspondingauthor{Seiji Fujimoto}
\email{seiji.fujimoto@utoronto.ca}
\author[0000-0001-7201-5066]{Seiji Fujimoto}
\affiliation{David A. Dunlap Department of Astronomy and Astrophysics, University of Toronto, 50 St. George Street, Toronto, Ontario, M5S 3H4, Canada}
\affiliation{Dunlap Institute for Astronomy and Astrophysics, 50 St. George Street, Toronto, Ontario, M5S 3H4, Canada}

\author[0000-0002-9382-9832]{Andreas L. Faisst}
\affiliation{IPAC, California Institute of Technology, 1200 E. California Blvd. Pasadena, CA 91125, USA}

\author[0000-0002-0498-5041]{Akiyoshi Tsujita}
\affiliation{Institute of Astronomy, Graduate School of Science, The University of Tokyo, 2-21-1 Osawa, Mitaka, Tokyo 181-0015, Japan}

\author[0000-0003-1041-7865]{Mahsa Kohandel}
\affiliation{Scuola Normale Superiore, Piazza dei Cavalieri 7, I-56126 Pisa, Italy}

\author[0000-0001-7457-4371]{Lilian L. Lee}
\affiliation{Max-Planck-Institute f\"ur extratarrestrische Physik, Giessenbachstrasse 1, 85748 Garching, Germany}

\author[0000-0003-4891-0794]{Hannah \"Ubler}
\affiliation{Max-Planck-Institute f\"ur extratarrestrische Physik, Giessenbachstrasse 1, 85748 Garching, Germany}

\author[0000-0002-8858-6784]{Federica Loiacono}
\affiliation{INAF – Osservatorio di Astrofisica e Scienza dello Spazio di Bologna, Via Gobetti 93/3, 40129 Bologna, Italy}

\author[0000-0002-7755-8649]{Negin Nezhad}
\affiliation{Department of Physics and Astronomy, University of California, Riverside, 900 University Ave, Riverside, CA 92521, USA}

\author[0000-0002-7129-5761]{Andrea Pallottini}
\affiliation{Dipartimento di Fisica ``Enrico Fermi'', Universit\'{a} di Pisa, Largo Bruno Pontecorvo 3, Pisa I-56127, Italy}

\author[0000-0002-6290-3198]{Manuel Aravena}
\affiliation{Instituto de Estudios Astrof\'{\i}sicos, Facultad de Ingenier\'{\i}a y Ciencias, Universidad Diego Portales, Av. Ej\'ercito 441, Santiago 8370191, Chile}
\affiliation{Millenium Nucleus for Galaxies (MINGAL), Av. Ej\'ercito 441, Santiago 8370191, Chile}

\author[0000-0002-9508-3667]{Roberto J. Assef}
\affiliation{Instituto de Estudios Astrof\'isicos, Facultad de Ingenier\'ia y Ciencias, Universidad Diego Portales, Av. Ej\'ercito Libertador 441, Santiago 8370191, Chile}

\author[0000-0003-4569-2285]{Andrew J. Battisti}
\affiliation{International Centre for Radio Astronomy Research (ICRAR), The University of Western Australia, M468, 35 Stirling Highway, Crawley, WA 6009, Australia}
\affiliation{Research School of Astronomy and Astrophysics, Australian National University, Cotter Road, Weston Creek, ACT 2611, Australia}

\author[0000-0002-3915-2015]{Matthieu B\'ethermin}
\affil{Universit\'e de Strasbourg, CNRS, Observatoire astronomique de Strasbourg, UMR 7550, 67000 Strasbourg, France}

\author[0000-0003-0946-6176]{Médéric Boquien}
\affiliation{Université Côte d'Azur, Observatoire de la Côte d'Azur, CNRS, Laboratoire Lagrange, 06000, Nice, France}

\author[0000-0001-9759-4797]{Elisabete da Cunha} 
\affiliation{International Centre for Radio Astronomy Research (ICRAR), The University of Western Australia, M468, 35 Stirling Highway, Crawley, WA 6009, Australia}

\author[0000-0002-9400-7312]{Andrea Ferrara}
\affil{Scuola Normale Superiore, Piazza dei Cavalieri 7, 50126 Pisa, Italy}

\author[0000-0002-3560-8599]{Maximilien Franco}
\affiliation{Université Paris-Saclay, Université Paris Cité, CEA, CNRS, AIM, 91191 Gif-sur-Yvette, France}

\author[0000-0002-9122-1700]{Michele Ginolfi}
\affiliation{Universit\`a di Firenze, Dipartimento di Fisica e Astronomia, via G. Sansone 1, 50019 Sesto Fiorentino, Florence, Italy}
\affiliation{INAF -- Arcetri Astrophysical Observatory, Largo E. Fermi 5, I-50125, Florence, Italy}

\author[0009-0003-3097-6733]{Ali Hadi}
\affiliation{Department of Physics and Astronomy, University of California, Riverside, 900 University Ave, Riverside, CA 92521, USA}

\author[0009-0006-3071-7143]{Aryana Haghjoo}
\affiliation{Department of Physics \& Astronomy, University of California, Riverside, 900 University Ave., Riverside, CA 92521, USA}

\author[0000-0002-2775-0595]{Rodrigo Herrera-Camus}
\affiliation{Departamento de Astronomía, Universidad de Concepción, Barrio Universitario, Concepción, Chile}
\affiliation{Millenium Nucleus for Galaxies (MINGAL), Av. Ej\'ercito 441, Santiago 8370191, Chile}

\author[0000-0003-4268-0393]{Hanae Inami}
\affiliation{Hiroshima Astrophysical Science Center, Hiroshima University, 1-3-1 Kagamiyama, Higashi-Hiroshima, Hiroshima 739-8526, Japan}

\author[0000-0002-6610-2048]{Anton M. Koekemoer}
\affiliation{Space Telescope Science Institute, 3700 San Martin Drive, Baltimore, MD 21218, USA} 

\author[0000-0002-1428-7036]{Brian C.\ Lemaux}
\affiliation{Gemini Observatory, NSF NOIRLab, 670 N. A'ohoku Place, Hilo, Hawai'i, 96720, USA}
\affiliation{Department of Physics and Astronomy, University of California, Davis, One Shields Ave., Davis, CA 95616, USA}

\author{Yuan Li}
\affiliation{Department of Physics and Astronomy and George P. and Cynthia Woods Mitchell Institute for Fundamental Physics and Astronomy, Texas A\&M University, 4242}

\author[0009-0004-1270-2373]{Lun-Jun Liu}
\affiliation{Physics Department, California Institute of Technology, 1200 E. California Blvd., Pasadena, CA, 91125 USA}

\author[0000-0002-8136-8127]{Juan Molina}
\affiliation{Instituto de F\'{i}sica y Astronom\'{i}a, Universidad de Valpara\'{i}so, Avda. Gran Breta\~{n}a 1111, Valpara\'{i}so, Chile}
\affiliation{Millenium Nucleus for Galaxies (MINGAL), Avda. Gran Breta\~{n}a 1111, Valpara\'{i}so, Chile}

\author[0000-0001-6652-1069]{Ambra Nanni}
\affiliation{National Centre for Nuclear Research, ul. Pasteura 7, 02-093 Warsaw, Poland}
\affiliation{INAF - Osservatorio astronomico d'Abruzzo, Via Maggini SNC, 64100, Teramo, Italy}

\author[0000-0002-7412-647X]{Francesca Pozzi}
\affiliation{University of Bologna – Department of Physics and Astronomy “Augusto Righi" (DIFA), Via Gobetti 93/2, 40129 Bologna, Italy}
\affiliation{INAF-Osservatorio di Astrofisica e Scienza dello Spazio, Via Gobetti 93/3, 40129, Bologna, Italy}

\author[0000-0003-1682-1148]{Monica Relano}
\affiliation{Dept. Física Te\'{o}rica y del Cosmos, Campus de Fuentenueva, Edificio Mecenas, Universidad de Granada, E-18071, Granada, Spain}
\affiliation{Instituto Universitario Carlos I de Física Te\'{o}rica y Computacional, Universidad de Granada, 18071, Granada, Spain}

\author[0000-0002-9948-3916]{Michael Romano}
\affiliation{Max-Planck-Institut für Radioastronomie, Auf dem Hügel 69, 53121 Bonn, Germany}
\affiliation{INAF - Osservatorio Astronomico di Padova, Vicolo dell'Osservatorio 5, I-35122 Padova, Italy}

\author[0000-0002-1233-9998]{David B. Sanders}
\affiliation{Institute for Astronomy, University of Hawaii, 2680 Woodlawn Drive, Honolulu, HI 96822, USA}

\author[0000-0003-4264-3381]{Natascha M. F\"orster Schreiber}
\affiliation{Max-Planck-Institute f\"ur extratarrestrische Physik, Giessenbachstrasse 1, D-85748 Garching, Germany}

\author[0000-0002-0000-6977]{John Silverman}
\affiliation{Kavli Institute for the Physics and Mathematics of the Universe, The University of Tokyo, Kashiwa, Japan 277-8583 (Kavli IPMU, WPI)}
\affiliation{Department of Astronomy, School of Science, The University of Tokyo, 7-3-1 Hongo, Bunkyo, Tokyo 113-0033, Japan}

\author[0000-0003-3256-5615]{Justin Spilker}
\affiliation{Department of Physics and Astronomy and George P. and Cynthia Woods Mitchell Institute for Fundamental Physics and Astronomy, Texas A\&M University, 4242}

\author[0000-0002-7919-245X]{Kseniia Telikova}
\affiliation{Instituto de Estudios Astrof\'isicos, Facultad de Ingenier\'ia y Ciencias, Universidad Diego Portales, Av. Ej\'ercito Libertador 441, Santiago 8370191, Chile}

\author[0000-0002-5877-379X]{Vicente Villanueva}
\affiliation{Departamento de Astronom{\'i}a, Universidad de Concepci{\'o}n, Barrio Universitario, Concepci{\'o}n, Chile}

\author[0000-0002-3258-3672]{Livia Vallini}
\affiliation{INAF – Osservatorio di Astrofisica e Scienza dello Spazio di Bologna, Via Gobetti 93/3, 40129 Bologna, Italy}

\author[0000-0002-7964-6749]{Wuji Wang}
\affiliation{Caltech/IPAC, 1200 E. California Blvd. Pasadena, CA 91125, USA}

\author[0000-0002-2318-301X]{Giovanni Zamorani}
\affiliation{INAF – Osservatorio di Astrofisica e Scienza dello Spazio di Bologna, Via Gobetti 93/3, 40129 Bologna, Italy}


\def\apj{ApJ}%
\def\apjl{ApJL}%
\def\apjs{ApJS}%

\def\rme{\rm e}
\def\rmstar{\rm star}
\def\rmFIR{\rm FIR}
\def\itHubble{\it Hubble}
\def\rmyr{\rm yr}

\begin{abstract}
We present a statistical study of spatially resolved chemical enrichment in 18 main-sequence galaxies at $z=4$--6, observed with \jwst/NIRSpec IFU as part of the ALPINE–CRISTAL–\jwst\ survey. 
Performing an optimized reduction and calibration procedure, including local background subtraction, light-leakage masking, stripe removal, and astrometry refinement, we achieve robust emission-line mapping on kiloparsec scales. 
Although line-ratio distributions vary across galaxies in our sample, 
we generally find mild central enhancements in [O\,\textsc{iii}]/H$\beta$, [O\,\textsc{ii}]/[O\,\textsc{iii}], [S\,\textsc{ii}]$_{6732}$/[S\,\textsc{ii}]$_{6718}$, H$\alpha$/H$\beta$, and $L_{\rm H\alpha}/L_{\rm UV}$, consistent with elevated electron density, dust obscuration, and bursty star formation accompanied by reduced metallicity and ionization parameter. 
These features point to inside-out growth fueled by recent inflows of pristine gas. 
Nevertheless, the median metallicity gradient is nearly flat over a few kpc scale, $\Delta \log({\rm O/H}) = 0.02 \pm 0.01$ dex kpc$^{-1}$, implying efficient chemical mixing through inflows, outflows, and mergers. 
From pixel-by-pixel stellar and emission-line characterizations, we further investigate the resolved Fundamental Metallicity Relation (rFMR). 
Metallicity is described by a fundamental plane with stellar mass and SFR surface densities, but with a stronger dependence on $\Sigma_{\rm SFR}$ than seen in local galaxies. 
Our results indicate that the regulatory processes linking star formation, gas flows, and metal enrichment were already vigorous $\sim$1 Gyr after the Big Bang, producing the nearly flat metallicity gradient and a stronger coupling between star formation and metallicity than observed in evolved systems in the local universe.
\end{abstract}

\section{Introduction}
\label{sec:intro} 

Understanding the chemical enrichment in the early universe is fundamental shed light on the processes of galaxy formation and evolution \citep[see reviews e.g.,][]{maiolino2019}. Gas-phase metallicity ($Z_{\rm gas}$) serves as a direct probe of chemical enrichment, quantifying the evolutionary stage of galaxies and revealing their star formation histories. Metals in the interstellar medium are the end products of stellar nucleosynthesis and are dispersed through various feedback mechanisms, such as supernova explosions and active galactic nuclei (AGN). These metals not only provide a record of past stellar nucleosynthesis but also play critical roles in the cooling of gas, which regulates future star formation and shapes the observed properties of nebular emission lines. 
Therefore, investigating $Z_{\rm gas}$ is essential for building a comprehensive understanding of galaxy evolution in the early universe.

Spatially-resolved measurements of $Z_{\rm gas}$ offer unique insights into the physical processes driving galaxy evolution. Such measurements enable the study of metallicity gradients and their connection to fundamental scaling relations, such as the mass-metallicity relation (MZR; e.g., \citealt{tremonti2004}) and the fundamental metallicity relation (FMR; e.g., \citealt{mannucci2010}). These relations are key to understanding the interplay between star formation, feedback, and gas accretion across different regions of galaxies. By combining spatially-resolved metallicity maps with kinematic information, we can also assess the impact of outflows, inflows, and mergers on the chemical enrichment of galaxies \citep[e.g.,][]{cresci2010, swinbank2012b, wuyts2016, wang2017}. 

In the local universe, integral field spectroscopy surveys have provided extensive datasets, such as MaNGA \citep[e.g.,][]{bundy2015}, SAMI \citep[e.g.,][]{croom2012}, and CALIFA \citep[e.g.,][]{sanchez2012}. These studies revealed that low-mass galaxies exhibit relatively flat metallicity gradients, while higher-mass galaxies tend to have steeper negative gradients, indicating more pronounced central enrichment \citep[e.g.,][]{belfiore2017, poetrodjojo2018}. Additionally, the spatially resolved mass-metallicity relation (rMZR) was established, showing that regions within galaxies follow a similar trend to the global mass-metallicity relation \citep[e.g.,][]{rosales-ortega2012,barrera-ballesteros2016}.

At intermediate redshifts ($z\simeq1$--3), the challenge of decreased spatial resolution is mitigated by utilizing gravitational lensing and advanced instrumentation. Observations with MUSE, SINFONI, and KMOS enabled studies of lensed galaxies, achieving sub-kiloparsec resolution \citep{jones2010,jones2013,yuan2011}. These investigations confirmed the presence of the resolved star-forming main sequence (rSFMS) and rMZR at these epochs, suggesting that the fundamental processes governing star formation and chemical enrichment were already in place \citep{wang2016, curti2020}. For instance, the KLEVER survey, using KMOS, provided spatially resolved metallicity maps and gradients in galaxies at $1.2 < z < 2.5$ \citep{curti2020}. The results revealed that metallicity gradients at these redshifts are generally flatter compared to local galaxies, suggesting a more uniform metal distribution likely driven by efficient gas mixing due to stronger gas inflows and outflows. This highlights the dynamic interplay between star formation, feedback processes, and gas accretion in the distant universe. However, extending similar observations to even more distant galaxies has been challenging due to wavelength-dependent limitations, as well as constraints in spatial resolution and the associated sensitivity requirements.

The advent of \jwst/NIRSpec IFU \citep{jakobsen2022, boker2022} has revolutionized our ability to perform spatially resolved metallicity measurements at even higher redshifts. Recent NIRSpec IFU studies have provided initial insights into metallicity gradients and their connection to galaxy dynamics and feedback processes in several galaxies at $z>6$. For instance, \citet{venturi2024} analyzed three systems at $z=6$--8 and found that their gas-phase metallicity gradients are consistent with being flat, suggesting efficient gas mixing likely driven by mergers or intense stellar feedback. In contrast, \citet{marconcini2024a} investigated the spatially-resolved gas-phase metallicity distribution in a galaxy at $z=9.11$ and reported a gradient, indicating spatial variations in chemical enrichment possibly due to differing star formation histories or gas accretion events. However, due to the time-intensive nature of NIRSpec IFU observations, these studies remain limited in sample size and often focus on pre-selected, well-known targets at $z>6$, introducing potential biases in our understanding of early galaxy evolution.
This suggests the importance of applying the similar spatially-resolved measurements systematically for normal main-sequence galaxies. 

In this paper, we report initial results of spatially-resolved metallicity measurements from an extensive NIRSpec IFU survey targeting 18 star-forming galaxies at $z=$4--6, selected from the ALMA Large Programs ALPINE \citep{lefevre2020,faisst2020,bethermin2020} and CRISTAL \citep{herrera-camus2025}. These galaxies reside on the main sequence of star formation in the stellar mass range of $\simeq10^{9.5}$--$10^{10.5}\,M_{\odot}$ and offer a statistically significant sample to bridge the gap between the previous results at $z\simeq0$--3 \citep[e.g.,][]{swinbank2012b, molina2017}, and the recent initial findings at $z>6$ enabled by \jwst. By providing spatially resolved metallicity measurements for a representative sample at $z=4-6$, this study establishes a foundation for investigating the evolution of galaxies over cosmic time at resolutions of a few hundred pc, unprecedented at these epochs. 

The paper is structured as follows. Section~\ref{sec:data} describes the data reduction and processing. Section~\ref{sec:analysis} outlines the methods for measuring emission line fluxes, ratios, and metallicity. Section~\ref{sec:result} presents the results of the spatially-resolved measurements for the metallicity and scaling relations, and discusses their implications. 
Finally, Section~\ref{sec:summary} summarizes our findings. 
We note that this paper is complementary to the analysis presented by L.~Lee et al.\ (in prep.). 
While our study focuses on presenting the data, establishing a statistical reference for redshift evolutions of the metallicity gradients and the resolved FMR at $z=4$--6, 
Lee et al.\ explore the impact of different metallicity diagnostics and calibrations and place stronger emphasis on the connection with galaxy kinematics. 
The two studies are generally consistent, with only minor differences that can be explained by the different scientific goals and corresponding analysis choices 
(e.g., in multiple-component systems, where Lee et al.\ analyze the component with reliable kinematic measurements while we consider the full system). 
Together, these complementary efforts provide a more complete picture of chemical enrichment in the ALPINE--CRISTAL sample and strengthen the collaborative framework of the survey.

Throughout this paper, we assume a flat universe the latest constraints from Planck \citep{planck2020}; 
$\Omega_{\rm m} =$ 0.3111, 
$\Omega_\Lambda =$ 0.6889, 
and $H_0 =$ 67.66 km s$^{-1}$ Mpc$^{-1}$. 
We use a \cite{chabrier2003} initial mass function (IMF) and magnitudes in the AB system \citep{oke1983}.

\section{Data processing} 
\label{sec:data}

\subsection{Observations}
\label{sec:obs}

The details of the observations and their connection to previous programs are presented in A.~Faisst et al.\ (submitted), and here we provide a brief summary.
A total of 18 star-forming galaxies at $z \simeq 4$--6 were observed with \jwst/NIRSpec IFU as part of the GO Cycle~2 program (\#3045, PI: A. Faisst) during April--May 2024. The targets were selected from galaxies observed in ALMA large programs, including ALPINE \citep[e.g.,][]{lefevre2020, bethermin2020, faisst2020} and CRISTAL \citep[e.g.,][]{herrera-camus2025}, in the COSMOS field \citep{scoville2007}. 
All targets have deep, and moderately high-resolution ($\simeq0\farcs2$--$0\farcs5$) \cii~158~$\mu$m and dust continuum data obtained with ALMA Band~7. Apart from DC-417567, all targets also have \jwst/NIRCam imaging in F115W, F150W, F277W, and F444W from COSMOS-Web \citep[e.g.,][]{casey2023}, while the remaining DC-417567 is covered with imaging from COSMOS \hst\ ACS F814W \citep{koekemoer2007} and \hst\ WFC3 from CANDELS \citep{grogin2011,koekemoer2011}, including F160W.
Among the 17 targets with NIRCam data, 4 (DC-630594, DC-742174, VC-5100994794, VC-5101244930) were also observed with the F090W, F200W, F356W, and F410M filters, as part of the PRIMER program \citep[e.g.,][]{donnan2024}. These datasets were combined and aligned onto a common grid with a pixel resolution of $0\farcs03$ using a drizzling method. Detailed procedures for the reduction of the NIRCam and HST data are presented in \cite{franco2025}. In this study, we utilize these processed imaging resources when necessary.

\begin{figure*}[!htbp]
\begin{center}
\includegraphics[trim=0cm 0cm 0cm 0cm, clip, angle=0,width=1\textwidth]{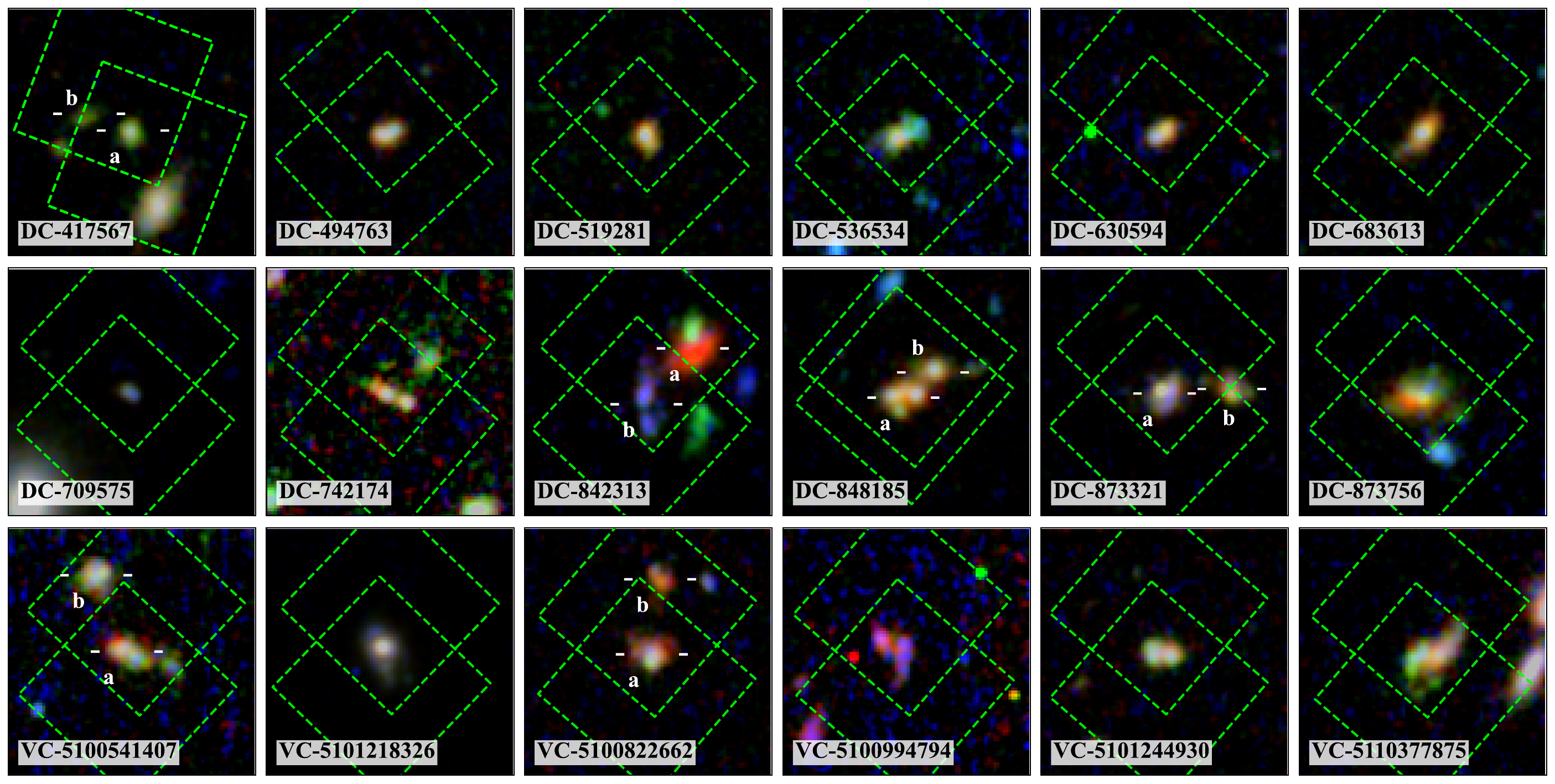}
\end{center}
\vspace{-0.2cm}
 \caption{
 Overview of the ALPINE–CRISTAL \jwst/NIRSpec IFU sample. 
Each panel shows a $5''\times5''$ RGB composite (F150W, F277W, F444W from JWST/NIRCam, 
except DC-417567 which uses HST ACS F814W, and WFC3 F125W, and F160W). 
The green dashed box indicates the footprint of the NIRSpec IFU in the two-point sparse cycling dither pattern, 
which is designed to place the primary target within the overlapped region so that it receives full exposure coverage.
In several cases, galaxies show spatially distinct components at the same spectroscopic redshift within the IFU. 
These components are labeled ``a'' and ``b'' and their radial properties are analyzed separately (Section~\ref{sec:radial_grad}).
\label{fig:target}}
\end{figure*}

All targets were observed using the IFU configurations G235M/F170LP 
(1.7--3.1~$\mu$m, $R\sim1000$) and G395M/F290LP (2.9--5.1~$\mu$m, $R\sim1000$), 
except for DC-842313, which was observed only with G235M/F170LP. 
Deeper G395H/F290LP observations (2.9--5.1~$\mu$m, $R\sim2700$) for DC-842313 were conducted in a separate program 
(\#4265, PIs J.~Gonz\'alez-L\'opez \& M.~Aravena; see also \citealt{solimano2024b}) during the same cycle, and are included in the analysis presented here. 
For program \#3045, we adopted a 2-point large sparse-cycling dither pattern, which provides two independent sub-pixel samplings of the $0\farcs1$ IFU spaxels. 
Each target was observed with $\sim$30--60 groups per integration and 1--3 integrations per exposure, yielding total on-source exposure times of $\sim$500--7000~s. 
The exposure times were optimized for each galaxy to ensure robust detections of the key rest-frame optical emission lines with sufficient S/N, according to the source brightness.
Target acquisition (TA) was requested for DC-848185 that shows an extended, multi-component morphology (see Figure~\ref{fig:target}), and thus the accurate TA was required to surely enclose the extended structures.  
We employed WATA (Wide Aperture Target Acquisition) using a star with F140W magnitude $\sim$20.5~AB, located 18\arcsec\ from the IFU center.  
For the remaining galaxies, which are compact ($\lesssim$0.5\arcsec\ relative to the IFU field of view), no active TA was performed. 
In \#4265, a 9-point small cycling dither was used with $\sim$18 groups per integration and 9 integrations per exposure, yielding 11,948~s of on-source exposure. 
No active TA was performed. Neither program included off-center background or light-leakage calibration exposures, and we therefore addressed these effects in post-processing (Section~\ref{sec:reduction}).
Figure~\ref{fig:target} shows RGB images of our 18 targets with the NIRSpec IFU FoVs overlayed.
Observation details for each target are summarized in Table~\ref{tab:obslog}. 
\begin{table*}[!htbp]
\begin{center}
\setlength{\tabcolsep}{3pt}
\caption{Summary of our target \& observation setup with IFU noise estimates}
\vspace{-0.2cm}
\label{tab:obslog}
\begin{tabular}{cccccccccc}
\hline \hline
Target Name          & $z$    & R.A        & Dec.   & Obs.~Date & Obs.~Mode & Exposure [s] &  $\sigma_{\rm G235M}$ & $\sigma_{\rm G395M/H}$ \\ 
(1) & (2) & \multicolumn{2}{c}{(3)} & (4) & (5) & (6) & (7) & (8) \\
\hline
DC-417567/C-10  & 5.6700 & 150.5170 & 1.9289 & 2024-05-26 & G235M/G395M & 7119/7236   & 11.9 & 3.7 \\
DC-494763/C-20  & 5.2337 & 150.0213 & 2.0534 & 2024-05-20 & G235M/G395M & 3939/7236   & 15.5 & 3.8 \\
DC-519281/C-09  & 5.5759 & 149.7537 & 2.0910 & 2024-04-25 & G235M/G395M & 1021/3764   & 31.5 & 5.6 \\
DC-536534/C-03$^{\sharp}$  & 5.6886 & 149.9718 & 2.1182 & 2024-04-27 & G235M/G395M & 1080/3676   & 30.3 & 5.7 \\
DC-630594/C-11  & 4.4403 & 150.1358 & 2.2579 & 2024-05-21 & G235M/G395M & 3852/7236   & 17.2 & 4.3 \\
DC-683613/C-05  & 5.5420 & 150.0393 & 2.3372 & 2024-05-21 & G235M/G395M & 2101/7236   & 22.8 & 4.2 \\
DC-709575/C-14  & 4.4121 & 149.9461 & 2.3758 & 2024-04-27 & G235M/G395M & 7236/7236   & 10.4 & 3.8 \\
DC-742174/C-17  & 5.6360 & 150.1630 & 2.4256 & 2024-05-23 & G235M/G395M & 7236/7236   & 10.0 & 3.5 \\
DC-842313/C-01$^{\sharp}$  & 4.5537 & 150.2272 & 2.5764 & 2024-04-25 & G235M/G395H$^{\dagger}$ & 7236/11948$^{\dagger}$  & 21.1 & 2.1$^{\dagger}$ \\
DC-848185/C-02  & 5.2931 & 150.0896 & 2.5864 & 2024-05-17 & G235M/G395M & 1167/4902   & 29.0 & 5.0 \\
DC-873321/C-07  & 5.1542 & 150.0169 & 2.6266 & 2024-04-26 & G235M/G395M & 1401/7003   & 24.3 & 4.3 \\
DC-873756/C-24  & 4.5457 & 150.0113 & 2.6278 & 2024-04-25 & G235M/G395M & 408/467     & 92.8 & 27.1 \\
VC-5100541407/C-06 & 4.5630 & 150.2538 & 1.8094 & 2024-04-24 & G235M/G395M & 759/1459   & 42.1 & 7.8 \\
VC-5100822662/C-04 & 4.5205 & 149.7413 & 2.0809 & 2024-05-17 & G235M/G395M & 1576/5340  & 20.3 & 4.5 \\
VC-5100994794/C-13 & 4.5802 & 150.1714 & 2.2873 & 2024-05-24 & G235M/G395M & 2743/7236  & 17.7 & 3.5 \\
VC-5101218326/C-25 & 4.5739 & 150.3021 & 2.3146 & 2024-04-26 & G235M/G395M & 467/584    & 86.6 & 30.6 \\
VC-5101244930/C-15 & 4.5803 & 150.1986 & 2.3006 & 2024-05-26 & G235M/G395M & 1838/7236  & 19.0 & 3.7 \\
VC-5110377875 & 4.5505 & 150.3848 & 2.4084 & 2024-05-26 & G235M/G395M & 992/2334   & 33.6 & 7.4 \\
\hline
\end{tabular}
\vspace{-0.4cm}
\end{center}
\tablecomments{
(1) Target name / ID from ALPINE \citep{lefevre2020} / CRISTAL \citep{herrera-camus2025} surveys. 
(2) Spectroscopic redshift from ALMA \cii\ \citep{faisst2020}. 
(3) Source coordinates (deg). (4) \jwst\ observation date. (5) NIRSpec grating. (6) Exposure times. 
(7,8) Standard deviation of our reduced IFU data cube ($10^{-22}$ erg s$^{-1}$ cm$^{-2}{\rm \AA}^{-1}$ per pixel) at $\sim2.4\mu$m and $\sim4.1\mu$m for G235M and G395M/G395H, respectively. 
}
$\sharp$ The presence of AGNs has been reported in these systems from the broad line detection (W.~Ren et al. submitted) or from emission line
diagnostics \citep{solimano2025}. \\
$\dagger$ The G395H data is taken in another relevant program of PID~4265 (PI: J.~Gonz\'alez-Lopez; see \citealt{solimano2025}), while the data reduction and calibration is reprocessed in the same manner as other data in this paper. 
\end{table*}
Additional detail on the survey will be presented in  A.~Faisst et al. (submitted). 

\subsection{Data reduction}
\label{sec:reduction}

We reduced the NIRSpec/IFU raw data using the STScI pipeline\footnote{\url{https://github.com/spacetelescope/jwst}} (ver~1.16.0, \citealt{bushouse2024}) with the CRDS context \texttt{jwst\_1298.pmap}, generally following the procedure developed by the ERS TEMPLATES team (\#1355; PIs J.~Rigby \& J.~Vieira). 
A detailed description for the procedure is provided in \cite{rigby2023} (see also \citealt{welch2023, birkin2023}), with associated data reduction code publicly available\footnote{\url{https://github.com/JWST-Templates/Notebooks/blob/main/nirspec\_ifu\_cookbook.ipynb}}. 
Through checking output data cubes, we also implement several additional steps to improve the final data quality. In this subsection, we explain the general reduction and calibration procedure, and we address our additional implementations in the next subsection. 

The raw data were processed through the standard three stages of the \jwst\ pipeline: 
Stage~1 handles detector-level corrections, such as bias subtraction and jump detection; 
Stage~2 applies calibrations, including flat-fielding, WCS assignment, wavelength calibration;
and Stage~3 constructs 3D data cubes from the calibrated 2D exposures.
After Stage~1, 
we applied a severe DQ flagging using the DQ frame\footnote{\url{https://jwst-pipeline.readthedocs.io/en/latest/jwst/references\_general/references\_general.html\#data-quality-flags}} to the Level~1 products: we flagged all pixels apart from DQ $=$ 0 (Good) or 4 (Jump detected during exposure) in the following procedure. 
We also applied \texttt{NSClean} package \citep{rauscher2023} to the Level~1 products, which mitigates systematic vertical pattern noise (1/f) and snowballs, improving the quality of the resulting calibrated detector images.

In Stage~2, since the off-target background exposures were not taken in our programs, the background subtraction was not applied inside the pipeline. Instead, we performed the background subtraction as a post-process (Section \ref{sec:bkg}).
Although the pixels severely affected by the light leakages from known failed open shutters are already flagged (DQ $=$ 536870912) in the above DQ flagging, additional shutters may open intermittently during every exposure. 
We thus inspected all exposure frames and made manual masks for the light leakage from the intermittent MSA failed open shutters in each frame before Stage~3. In Appendix \ref{sec:appendix_mask}, we show an example of the light leakages we mask.  

In Stage 3, although several previous NIRSpec IFU studies \citep{cresci2023, marshall2023,perna2023,wwang2025} have reported that the default outlier detection during Stage~3 can lead to false outlier detections corresponding to bright sources in dithered exposures, we find
that spurious outlier detections do not occur in our data, likely because our targets are much fainter than those in previous studies such as the luminous quasars (see also e.g., \citealt{vanzella2023, fujimoto2024}) and the improvements of the internal algorithms in the pipeline, compared to the earlier versions (e.g., version 1.12.5 as used in \citealt{fujimoto2024}). 
We thus ran the outlier detection step with the default parameters.
Because the 2-point sparse cycling dithering employed for most of our observations provides limited sub-pixel sampling, we adopt the nominal pixel scale of $0\farcs1$ for the cube building in Stage~3. 

\subsection{Additional steps outside pipeline}
\label{sec:post}

Upon verifying the data cubes generated from Stage~3, we identified several residual issues requiring correction: systematic stripes in collapsed continuum maps, underestimated errors, and astrometry offset. Below, we describe the additional steps taken outside the pipeline to address these issues, in addition to the background subtraction and an absolute flux comparison between NIRSpec IFU and NIRCam.

\subsubsection{Background subtraction}
\label{sec:bkg}

\begin{figure}[!htbp]
\begin{center}
\includegraphics[trim=0cm 0cm 0cm 0cm, clip, angle=0,width=0.5\textwidth]{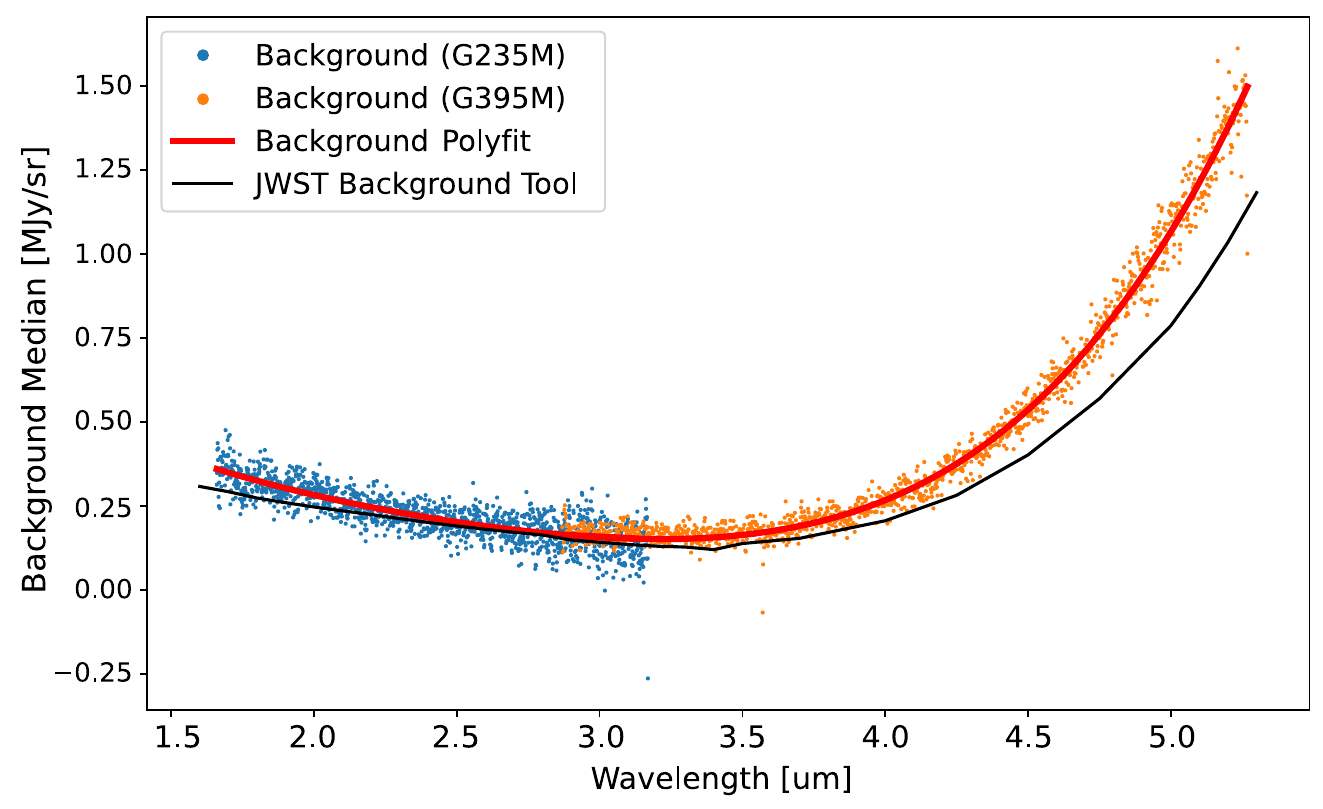}
\end{center}
\vspace{-0.2cm}
 \caption{
Example of the local background estimated for DC-630594. 
The color dots denote the median at the sky area in each channel of the IFU cubes. 
The red and black curves indicate the best-fit polynomial function and the predicted value at the target coordinate and observation date with \jwst\ background tool, respectively. 
Among our targets, these two curves are generally consistent within $\lesssim10\%$ at $\simeq$~2--4$\mu$m, while the gaps increase to $\simeq$~20--30\% at shorter and longer wavelengths. 
We adopt the best-fit polynomial function to subtract the local background in our IFU cubes. 
\label{fig:bkg}}
\end{figure}

Since off-target background exposures were not taken in our programs, we estimated the local background within each cube and channel for background subtraction, utilizing the large sky area covered by the 2-point sparse cycling dithering. 
For the background measurement, we calculated the median value in each channel, applying a sigma-clipping procedure (5$\sigma$). A 4th-order polynomial fit was then applied to the median values as a function of wavelength, and the background was subtracted using the best-fit function. 
Figure~\ref{fig:bkg} presents an example of our background estimate compared to the prediction from the \jwst\ background tool, based on the target coordinates and observation date.

We find that our estimate is generally consistent with the tool's prediction to within $\lesssim10\%$ at $\simeq$2--4~$\mu$m. However, gaps increase to $\simeq$ 20--30\% at shorter and longer wavelengths. This highlights the need for caution when relying on the \jwst\ background tool for background estimation in these wavelength regimes. 

\subsubsection{Stripes}
\label{sec:stripe}

\begin{figure}[!htbp]
\begin{center}
\includegraphics[trim=0cm 0cm 0cm 0cm, clip, angle=0,width=0.5\textwidth]{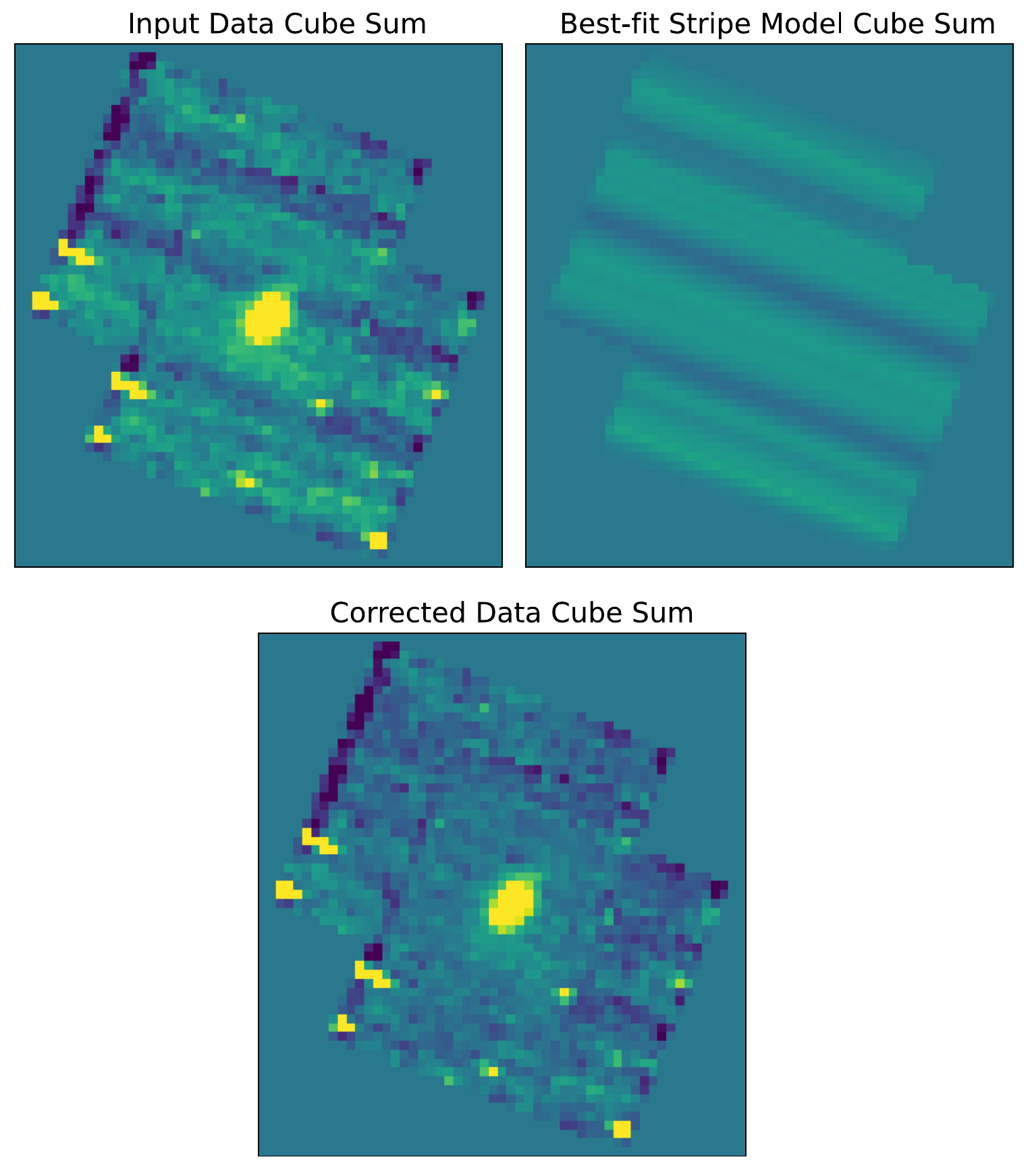}
\end{center}
\vspace{-0.2cm}
 \caption{
Systematic stripes emerged in the collapsed continuum map among our IFU cubes. Here we highlight DC-630594 as an example. 
From the top left, top right, and bottom panels, we show the continuum maps collapsing all channels of the G235M cube, the best-fit empirical stripe model cube (see Section~\ref{sec:stripe}), and the residual cube by subtracting the best-fit empirical stripe model from the observed cube. 
We use the residual cube in our analyses. 
\label{fig:stripe}}
\end{figure}

When creating continuum maps by collapsing all cube channels, we find the emergence of a systematic stripe pattern. The left panel of Figure~\ref{fig:stripe} presents an example of this stripe pattern in a collapsed continuum map. 
This stripe pattern is consistently present in all G235M, G395M, and G395H data, with variations observed across different cubes. A similar feature has been reported in other IFU programs (\citealt{decarli2024}; E. Vanzella et al., private communication), suggesting that it is a common feature independent of the observation mode. 
The orientation of the stripes aligns with the position angle of the IFU, corresponding to the vertical direction in the detector frame. This observation indicates that the pattern might originate from residual vertical noise remaining after the application of \texttt{NSClean} (Section~\ref{sec:reduction}). 
While these stripes are not prominent in individual channel maps and therefore have negligible impact on emission line analyses, they warrant caution for studies relying on continuum maps generated by collapsing IFU cubes.

We corrected the stripe feature by empirically modeling it, following the method described in \cite{decarli2024}. Specifically, we computed the median value of pixels along a 1-pixel-wide column tilted to match the IFU's position angle\footnote{The position angle information was retrieved from the header using the keyword \texttt{PA\_APER}} in each channel. For the median calculation, we masked pixels associated with detected emission lines ($\geq5\sigma$, based on \oiii). 
During this process, we encountered issues in some rows due to an insufficient number of pixels available for the median calculation. In such cases, we excluded rows with fewer than 10 available pixels and subsequently filled these values by interpolating the median values from neighboring rows.
Although the stripe pattern was modeled using the tilted cube, we ensured that the modeled stripe cube was re-aligned to the original position angle of the observed cube before subtraction. This approach prevented any additional resampling of the observed data cube. 
The top right and bottom panels of Figure~\ref{fig:stripe} present an example of the continuum map generated by collapsing our stripe model cube and the residual map obtained after subtracting the modeled stripe pattern from the observed map, respectively.

\subsubsection{Rescaling the Error Array}
\label{sec:error}

The data cubes generated in Stage~3 of the NIRSpec pipeline include an error frame in the \texttt{ERR} extension, which accounts for several uncertainties, such as Poisson noise and detector readout noise. 
However, when we compared the 1$\sigma$ error values derived from the sky area in the science frame with the median values from the corresponding regions in the error frame, we found that the error frame values were systematically underestimated (see also \citealt{ubler2023}). This difference likely arises because the pipeline's error propagation does not account for flat-fielding inaccuracies, residual background noise, or effects of 1/f stripe noise.
To address this, we measured the ratio of the observed 1$\sigma$ errors to the error frame values in each channel. We found that this ratio was approximately constant across the wavelength range. To simplify the correction, we thus performed a constant fit to the ratio as a function of wavelength and applied the best-fit scaling factor to renormalize the error frame values.
The typical scaling factor is estimated to be $\sim$1.5--2.0. 
For subsequent analyses, such as line flux measurements, we use the renormalized error frame values for the corresponding pixels to ensure accurate uncertainty estimates.

\subsubsection{Astrometry}
\label{sec:astrometry}

\begin{figure}[!htbp]
\begin{center}
\includegraphics[trim=0cm 0cm 0cm 0cm, clip, angle=0,width=0.5\textwidth]{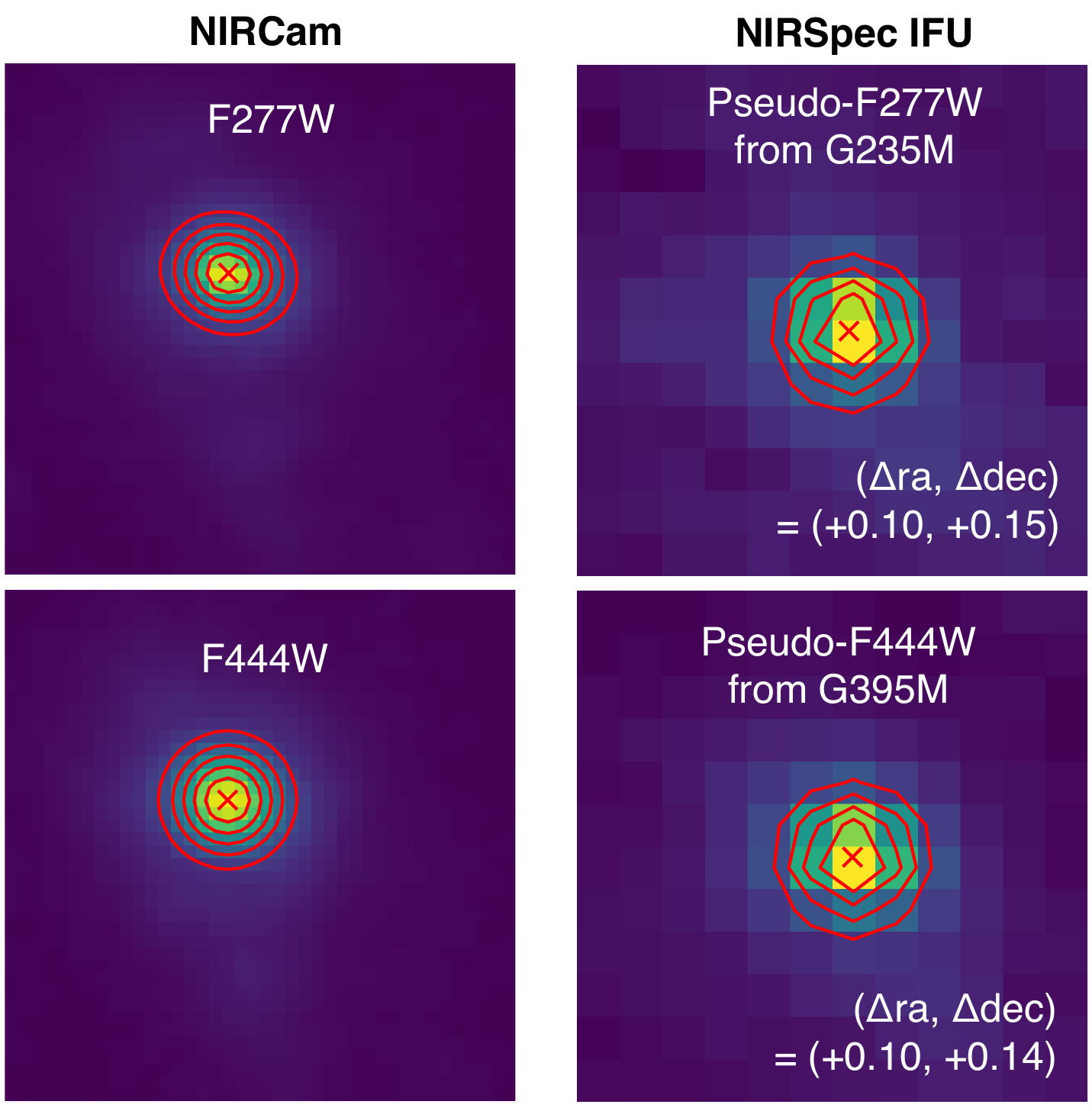}
\end{center}
\vspace{-0.2cm}
 \caption{
Astrometry correction procedure. Here we highlight VC-5101218326 as an example. 
The left panels present the actual NIRCam F277W and F444W maps, and the right panels represent the pseudo-F277W and F444W maps generated from the G235M and G395M cubes by convolving them with the NIRCam filter responses, respectively.  
We determine the source center by performing the 2D elliptical Gaussian fit to both images, evaluate the spatial offset between pseudo and actual F277W and F444W maps, and use the average $\Delta$R.A. and $\Delta$Decl. for the correction. 
The astrometry offset estimated by this procedure is summarized in the Appendix Table~\ref{tab:offset}. 
\label{fig:astrometry}}
\end{figure}

Regardless of the TA steps, NIRSpec IFU observations rely on blind pointing for the science exposures.\footnote{Even with the TA step, the final placement of the science target in NIRSpec IFU observations ultimately depends on blind pointing.}
In most cases, the targets are sufficiently contained within the IFU field of view (FoV; $3'' \times 3''$) using blind pointing, and this does not pose a critical issue. 
However, astrometry offsets on the scale of $\sim0\farcs1$ have been reported even when the WATA procedure is applied \citep{fujimoto2024}. Such offsets can become significant when comparing differential morphology or spatial offsets across multi-wavelength data, especially when combined with high-resolution datasets such as NIRCam images. To address this, we refined the astrometry of our IFU cubes through the following procedures, which are visually presented in Fig~\ref{fig:astrometry}.

First, we generated pseudo-F277W and F444W continuum images by convolving the G235M and G395M (or G395H) IFU cubes with the respective filter response curves of the NIRCam filters. For targets without available NIRCam data, we generated the pseudo-F277W image from the G235M cube and used the \hst\ WFC3 F160W image as the reference instead. Next, as shown in Figure~\ref{fig:astrometry}, we performed 2D elliptical Gaussian fits on both the actual NIRCam images (or \hst\ WFC3 F160W) and the pseudo-NIRCam F277W and F444W images to evaluate the positional offset of their peaks. Since the G235M and G395M observations were conducted consecutively for each target, we assumed that the astrometry offset should be consistent between the two, thus we averaged the offsets between the F277W and F444W peaks, and applied these values as corrections to refine the astrometry.
 For DC-842313, where G235M and G395H observations were conducted independently in different programs, the offset was evaluated separately.

In Appendix~\ref{sec:appendix_additional}, we summarizes the astrometry offsets for each target. We find that the offsets are typically $\sim0\farcs2$--$0\farcs3$, confirming that blind pointing generally ensures the targets remain within the IFU FoV. 
These offsets are consistent with the scales found also in previous studies \citep{arribas2024, jones2024a, jones2024b, lamperti2024, ubler2024, parlanti2025}. 
However, we caution that detailed morphological and spatial analyses comparing IFU data with other high-resolution datasets may still require additional care. We also observe minor variations in the offsets between G235M and G395M, typically $\simeq0\farcs00$--$0\farcs05$, indicating that this level of residual astrometric uncertainty remains in our IFU cubes.

\subsubsection{Flux Accuracy}
\label{sec:absolute_flux}

The pseudo-F277W and F444W continuum images generated from the NIRSpec IFU cubes can also be used to verify the flux calibration of NIRSpec IFU by comparing the flux with that observed in the actual NIRCam maps. Using a common $0\farcs5$-diameter aperture applied to both the pseudo and actual F277W and F444W NIRCam maps at the peak positions, we evaluate the flux difference as $|f_{\rm NIRSpec} - f_{\rm NIRCam}| / f_{\rm NIRSpec}$. 
In Appendix~\ref{sec:appendix_additional}, we summarizes the flux differences for each target. 
We find that the typical flux difference is close to zero, with a dispersion extending up to $\sim$30\%. This dispersion may arise from a combination of several uncertainties, such as the absolute flux calibration, background subtractions, and astrometry, in both NIRSpec and NIRCam. Given the overall consistency, we do not propagate this potential uncertainty from the absolute flux calibration of either instrument in our subsequent analyses. However, it is worth noting that this level of uncertainty may persist on an individual basis.

\subsubsection{PSF matching}
\label{sec:psf}

We match the point spread functions (PSFs) across all observations by making them consistent with the PSF of the F444W filter. The \jwst\ PSF is modeled using the \texttt{WebbPSF}\footnote{\url{https://github.com/spacetelescope/webbpsf}} package \citep{perrin2014}. For the NIRSpec/IFU data, the spectral cube is divided into wavelength bins of 0.5\,$\mu$m. A PSF model corresponding to the median wavelength of each bin is employed for PSF matching, ensuring representative adjustments. We use the PSF-matched IFU cubes for the following analyses. 

\subsection{Final products}
\label{sec:product}
\begin{figure*}[!htbp]
\begin{center}
\includegraphics[trim=0cm 0cm 0cm 0cm, clip, angle=0,width=1.\textwidth]{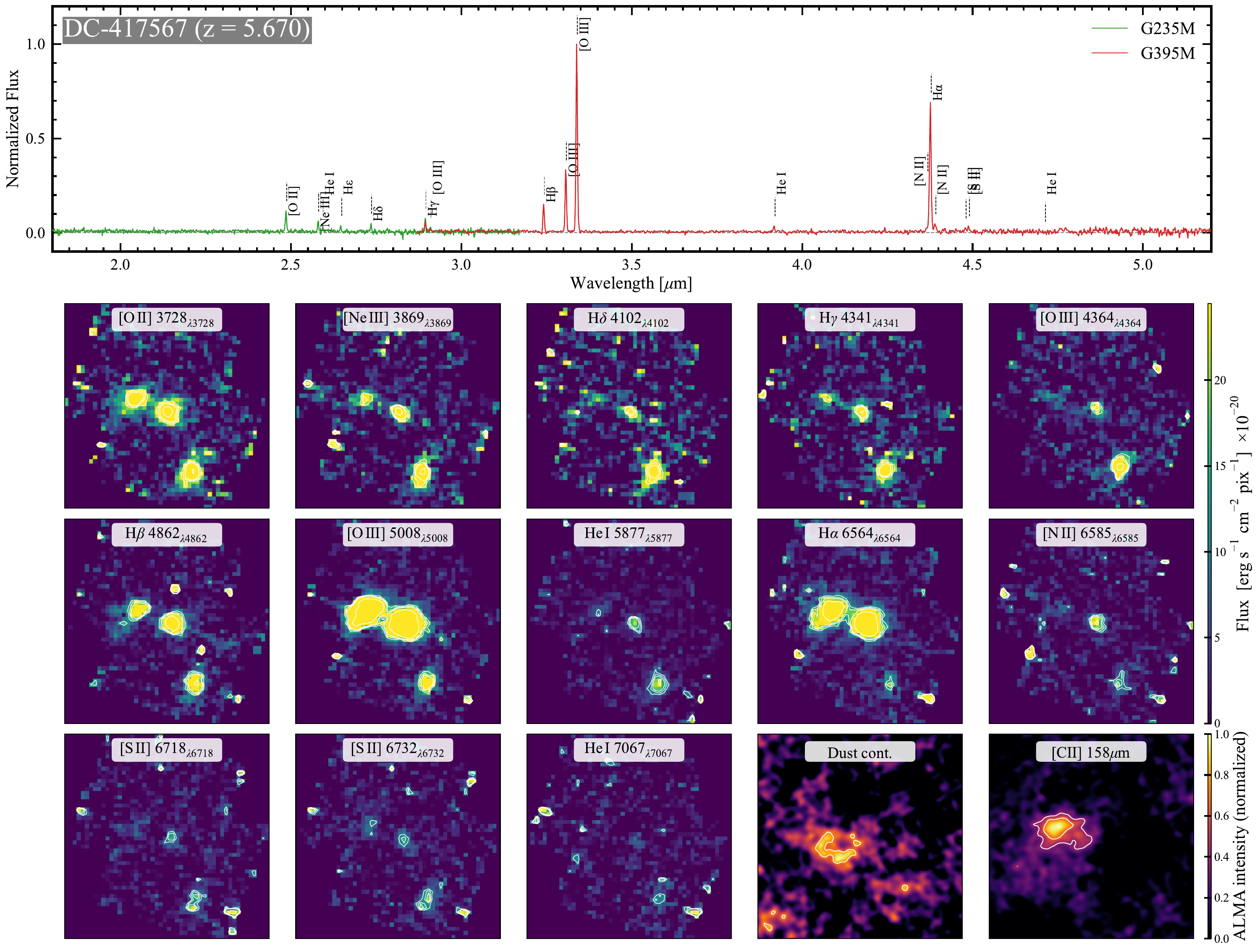}
\end{center}
\vspace{-0.4cm}
\caption{
Example of NIRSpec IFU products for the target DC-417567 at $z=5.67$. 
\textit{\textbf{Top:}} Extracted 1D spectra with a $0\farcs4$-radius aperture at the H$\alpha$ line peak from the G235M (green) and G395M (red) gratings, normalized to the maximum flux. 
Key emission lines are indicated by vertical dashed lines. 
\textit{\textbf{Bottom:}} Moment-0 maps ($5^{\prime\prime}\times5^{\prime\prime}$) for strong rest-optical emission lines observed with NIRSpec IFU, together with the ALMA dust continuum and [C\,\textsc{ii}] 158\,$\mu$m moment-0 maps. 
White contours show $3\sigma$, $5\sigma$, and $10\sigma$ significance levels. 
\label{fig:comb_example}}
\end{figure*}

Figure~\ref{fig:comb_example} presents an example of the 1D spectrum extracted from our reduced and calibrated IFU cubes.
The key emission lines are clearly detected, and their 2D maps are also highlighted, along with the NIRCam cutouts. In Appendix~\ref{sec:appendix_sample}, we summarize the 1D spectrum and 2D maps for all targets. These reduced and calibrated data cubes will be made publicly available online upon the acceptance of this paper.

\section{Data Analysis}
\label{sec:analysis}

\subsection{Line Flux Measurement and Line Maps}
\label{sec:flux}

To investigate spatially resolved properties, line fluxes are determined by performing Gaussian fitting on the 1D spectrum extracted from each pixel of the IFU cube. 
Our NIRSpec IFU data encompass key optical emission lines, including [O\,\textsc{ii}]$_{\lambda\lambda 3727,3730}$, H$\beta$,
[O\,\textsc{iii}]$_{\lambda 4363}$,
[O\,\textsc{iii}]$_{\lambda\lambda 4960,5008}$, H$\alpha$, [N\,\textsc{ii}]$_{\lambda\lambda 6548,6585}$, and [S\,\textsc{ii}]$_{\lambda\lambda 6716,6731}$. 
The \cii-based spectroscopic redshift and line width are used as priors for the fitting, while the line center and width remain as free parameters. 

Some closely spaced emission lines, such as \oiii+H$\beta$ and H$\alpha$+\nii, are fitted simultaneously, assuming the same line width, line redshift, and underlying continuum. 
During these simultaneous fits, the line ratios \nii$_{\lambda6548}$:\nii$_{\lambda6585}$ and \oiii$_{\lambda4960}$:\oiii$_{\lambda5008}$ are fixed to 1:2.94 and 1:2.98, respectively, consistent with the theoretical values in the low-density limit \citep[e.g.,][]{storey2000, osterbrock2006}. 

For each spaxel (spectral pixel), a Gaussian profile is first fitted to the H$\alpha$+\nii\ lines. A spaxel is considered to be a valid detection if the best-fit amplitude of H$\alpha$ exceeds three times its associated uncertainty. Based on these H$\alpha$ results, a detection mask is constructed and then applied uniformly to all other emission lines. To mitigate the contamination from noise spikes, only contiguous regions within the detection mask that contain at least nine pixels are retained for subsequent analysis; regions with fewer than nine connected pixels are discarded.
The total line flux in detected pixels is obtained by integrating the best-fit Gaussian, and we create line fluxes and associated error maps for each emission line. 
Compared to moment~0 maps generated by integrating over a wide velocity range, this approach is particularly advantageous for faint emission lines, where the signal may be easily embedded in noise fluctuations in the wide integrations.

\subsection{Gas-phase metallicity measurements}
\label{sec:zgas}

We measure gas-phase metallicities ($Z_{\rm gas}$) using the pixel-by-pixel based emission line fluxes obtained from the IFU cubes described in Section~\ref{sec:flux}. Prior to metallicity estimation, the emission line fluxes are corrected for dust attenuation ($A_{\rm V}$) using the Balmer decrement (H$\alpha$/H$\beta$) method in each pixel based on the \cite{calzetti2000} law. We assume the intrinsic H$\alpha$/H$\beta$ ratio of 2.86 under Case~B recombination conditions \citep{osterbrock2006}.

To achieve robust metallicity measurements with stable S/N on pixel-by-pixel basis, we utilize rest-frame optical strong emission lines, which benefit from extensive calibrations \citep[e.g.,][]{bian2018, curti2020b, nakajima2022b,sanders2023, sanders2025}. Specifically, we employ the R3 index ($\equiv$\oiii$_{\lambda5008}$/H$\beta$), which has an advantage of minimal impact from the dust correction uncertainty, while a downside is to yield double-valued metallicity solutions. 
We thus additionally use the O32 index ($\equiv$\oiii$_{\lambda5008}$/\oii$_{\lambda3727,3730}$) to resolve this degeneracy. The best-fit metallicity is derived from a simultaneous fitting of these line ratios to the latest calibration equations presented in \cite{sanders2025}, with uncertainties determined via a Monte Carlo approach by perturbing the line ratios with their measurement errors and repeating the process 500 times.

\begin{figure*}[!htbp]
\begin{center}
\includegraphics[trim=0cm 0cm 0cm 0cm, clip, angle=0,width=1.\textwidth]{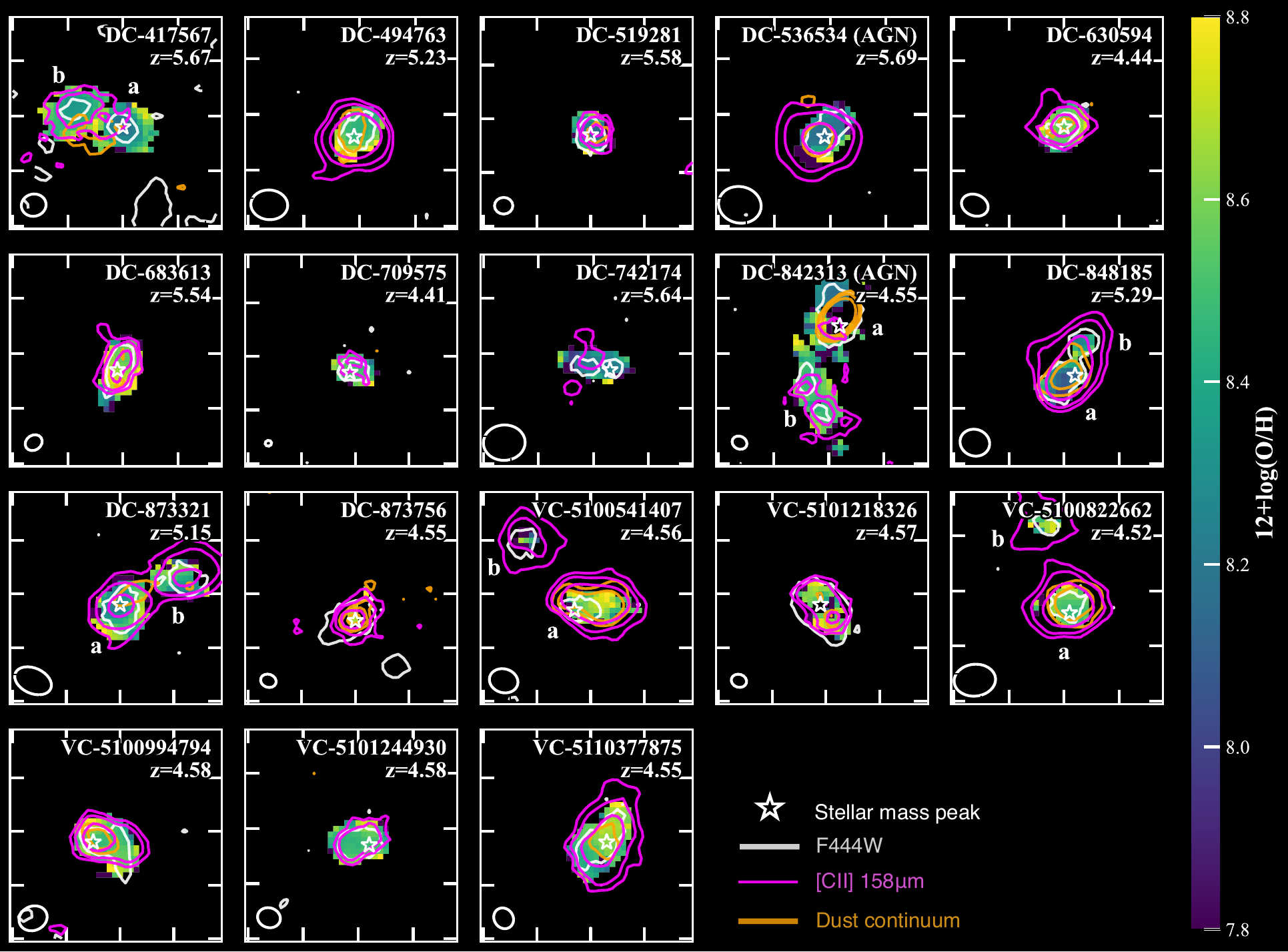}
\end{center}
\vspace{-0.2cm}
\caption{
Spatially resolved gas-phase metallicity maps for our galaxy sample, overlaid with rest-frame FIR and optical emission. 
Each panel shows a $4''\times4''$ cutout of $12+\log({\rm O/H})$, with the [C\,\textsc{ii}] 158\,$\mu$m emission 
(magenta contours; 3, 5, and 10$\sigma$ levels), the ALMA dust continuum 
(orange contours; 3, 5, and 10$\sigma$ levels), and the NIRCam F444W emission 
(white contours; 3$\sigma$ level) overplotted. 
White stars denote the stellar mass peak derived from pixel-by-pixel SED fitting. 
White open ellipses indicate the synthesized beam size of the ALMA data used for the \cii\ and dust continuum maps. 
DC-536534 and DC-842313 are identified as AGNs based on robust broad-line detections (W. Ren et al., submitted) and emission-line diagnostics \citep{solimano2025}, as indicated in their labels.
The labels of ``a'' and ``b'' are used in the same manner as Figure~\ref{fig:target}.
\label{fig:metal-fir}}
\end{figure*}

\begin{figure*}[!htbp]
\begin{center}
\includegraphics[trim=0cm 0cm 0cm 0cm, clip, angle=0,width=1.\textwidth]{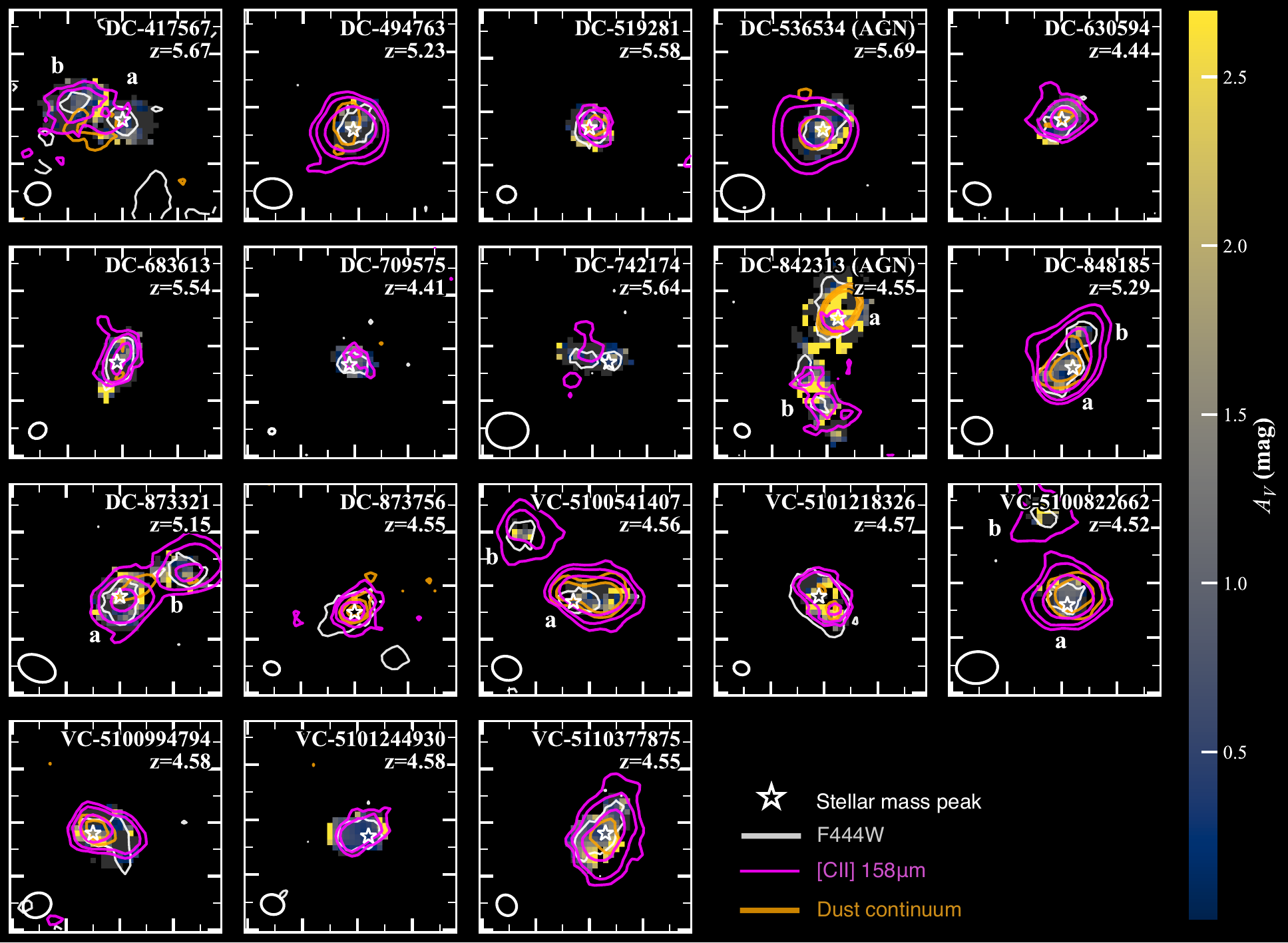}
\end{center}
\vspace{-0.2cm}
\caption{
Spatially resolved dust attenuation ($A_V$) maps overlaid with rest-frame FIR and optical emission for our sample.  
Each panel shows a $4''\times4''$ cutout of the $A_V$ distribution derived from the Balmer decrement using H$\alpha$/H$\beta$, and other symbols are the same as Figure~\ref{fig:metal-fir}.  
The labels of ``a'' and ``b'' are used in the same manner as Figure~\ref{fig:target}. 
\label{fig:av-fir}}
\end{figure*}

Figure~\ref{fig:metal-fir} and \ref{fig:av-fir} present the resulting metallicity and $A_{\rm V}$ distributions derived for our sample. 
We note that our metallicity measurements are generally consistent with those obtained in these targets using spatially-integrated emission line fluxes (A. Faisst et al. submitted). Furthermore, the auroral \oiii$\lambda4363$ line is well detected in the spatially-integrated photometry in five galaxies in our sample and also in the rest via stacking, allowing a direct temperature-based metallicity measurement. The metallicity derived via this direct method is confirmed to be consistent with our strong-line-based metallicity measurements within the uncertainty (A. Faisst et al., submitted).

We note that several of our targets have been reported as AGN candidates, either from broad Balmer line detections (W.~Ren et al., submitted) or from emission-line diagnostics \citep{solimano2025}. 
Among them, we classify DC-536534 and DC-842313 (dubbed ```C01-total'' in \citealt{solimano2025}) as AGNs based on the robust detection of a broad Balmer line (FWHM $\sim$3000~km~s$^{-1}$) and a significant offset from the star-forming regime in the BPT diagram, respectively. 
Because the presence of an AGN may bias strong-line metallicity calibrations, we exclude the AGN-like regions (via BPT) with a diameter of 2 pixels ($\approx0\farcs2$, equal to 1.3~kpc at $z=5$) in these two sources from the subsequent analyses (Sections~\ref{sec:radial_grad}, \ref{sec:Zgrad}, and \ref{sec:rFMR}), given that the PSF FWHM of the NIRSpec IFU at the relevant wavelengths is $\sim0\farcs15$--$0\farcs2$. 

\subsection{Pixel-by-pixel SED fitting}
\label{sec:sed}

To explore the spatially resolved relationships between nebular emission lines and stellar properties, such as stellar mass ($M_{\star}$), star formation rate (SFR), and stellar age ($t_{\rm age}$), we conducted pixel-by-pixel Spectral Energy Distribution (SED) fitting \citep[e.g.,][]{clara2023}. 
Detailed methodologies for the pixel-by-pixel SED fitting are provided in A.~Tsujita et al. (submitted).

In brief, we utilized the \texttt{Prospector} SED-fitting code \citep{leja2017,johnson2021}, combining \jwst/NIRCam broadband photometry with key emission lines detected by JWST/NIRSpec IFU observations, specifically H$\alpha$, H$\beta$, and \oiii$_{\lambda5008}$. We adopted a parametric delayed-$\tau$ star formation history (SFH) model, a \cite{chabrier2003} IMF, and a \cite{calzetti2000} attenuation curve with a power-law slope modification as a free parameter. 
Pixels were included in the analysis only if detections exceeded $3\sigma$ in at least two NIRCam filters after alignment and point-spread function (PSF) matching across all bands.

Our SED fitting results are generally consistent with previous studies targeting the same galaxies \citep[e.g.,][]{li2024,mitsuhashi2024}. However, we find some differences -- for instance, the $M_{\star}$ values are $\sim$0.4~dex (median) smaller than the estimates in \cite{li2024} and \cite{mitsuhashi2024}.  
While these previous NIRCam+ALMA based SED studies have advanced the characterizations of our targets, resolving the degeneracy between age and dust, we note that the limited available NIRCam broadband photometry, primarily restricted to the F277W and F444W bands, may lead to degeneracies between strong optical emission lines (e.g., H$\alpha$, [O{\sc iii}]) and the underlying continuum. Our approach, leveraging the NIRSpec IFU, directly resolves these degeneracies, enabling more robust derivations of physical properties. 
Note also that we do not include the ALMA data in our SED fitting analysis due to their lower angular resolution compared to the optical--NIR datasets, while we confirm the general consistency in the SED properties between the outputs with and without the ALMA photometry when performing the SED fitting for the integrated photometry (Tsujita et al. submitted).  

\section{Results \& Discussion}
\label{sec:result}

\subsection{2D Distributions of Metallicity and Dust Attenuation}
\label{sec:cii_dust}

First, we provide an overview of the spatial variations in chemical enrichment and dust attenuation together with the other physical properties of our targets. 
In Figures~\ref{fig:metal-fir} and~\ref{fig:av-fir},   
we present 2D maps of $Z_{\rm gas}$ and $A_{\rm V}$ (Section~\ref{sec:zgas}) and indicate the peak positions of the stellar mass distributions derived from pixel-by-pixel SED fitting (Section~\ref{sec:sed}), together with the contours of the [C\,\textsc{ii}] emission and the underlying dust continuum observed with ALMA. 

We find substantial diversity in the spatial distributions of both metallicity and dust attenuation across the sample. 
Some galaxies exhibit centrally enhanced metallicities, generally coinciding with their stellar mass peaks (e.g., DC-683613, VC-5110377875), whereas others show higher metallicities in their outskirts (e.g., DC-417567, DC-494763). 
A similar variety is seen in dust attenuation: in some systems the $A_{\rm V}$ peak aligns with the stellar mass center (e.g., DC-494763, DC-873321), while in others it is significantly offset (e.g., VC-5101244930, DC-5110377875). 
In most galaxies the [C\,\textsc{ii}] and dust continuum emission arise near the stellar mass peaks, with a few exceptions (DC-417567, VC-5100541407). 
Given the diverse distributions of metallicity and $A_{\rm V}$, the peaks of [C\,\textsc{ii}] and dust continuum often do not coincide with the regions of highest metallicity or dust attenuation (e.g., DC-494764, DC-848185, DC-873321). 
Such diversity likely reflects the combined influence of mergers, inflows, and outflows, which redistribute metals and dust, and create variations in local ISM conditions, ionization states, and recent bursts of star formation (see also \citealt{herrera-camus2025}).

Overall, the misalignments among metallicity, dust attenuation, FIR emission, and stellar mass reveal the complex internal physics of $z=4$--6 galaxies, highlighting the importance of multi-wavelength, spatially resolved studies to understand early galaxy evolution. 
To quantify these variations, we investigate radial gradients of key emission-line properties in Section~\ref{sec:radial_grad} and the resolved fundamental relation linking metallicity, stellar mass surface density ($\Sigma_{\star}$), and SFR surface density ($\Sigma_{\rm SFR}$) in Section~\ref{sec:rFMR}. 
A detailed analysis of the connection between the resolved galaxy properties and the FIR emission will be presented in forthcoming papers, including F.~Lopez et al.\ (in prep.) that investigate the relations between dust-to-gas and dust-to-stellar mass ratios and metallicity in a subsample, providing an independent perspective on the interplay between metals, dust, and gas at $z\sim4$--6.

\subsection{Radial gradients of emission properties at $z=4-6$}
\label{sec:radial_grad}

We also investigate average radial properties. By computing radially averaged fluxes, this approach effectively enhances the S/N, particularly in the galaxy outskirts, thus providing robust characterizations of the spatial variations of emission properties from the inner to outer regions of galaxies. 
To separately analyze companions, which may affect the radial measurements, we first create segmentation maps using the NIRCam F444W images\footnote{
The \hst/F160W is used for DC-417567 that is not observed with NIRCam. 
} by running \texttt{SExtractor} with default parameter settings\footnote{\url{https://github.com/astromatic/sextractor}}. 
These segmentation maps are then employed as masks to separate the central and companion galaxies. The segmented galaxies are labeled as ``\_a'' and ``\_b''. 
The flux measurements enclosed within each segmentation region are individually utilized in the following analyses. 

\begin{figure*}[!htbp]
\begin{center}
\includegraphics[trim=0cm 0cm 0cm 0cm, clip, angle=0,width=1.\textwidth]{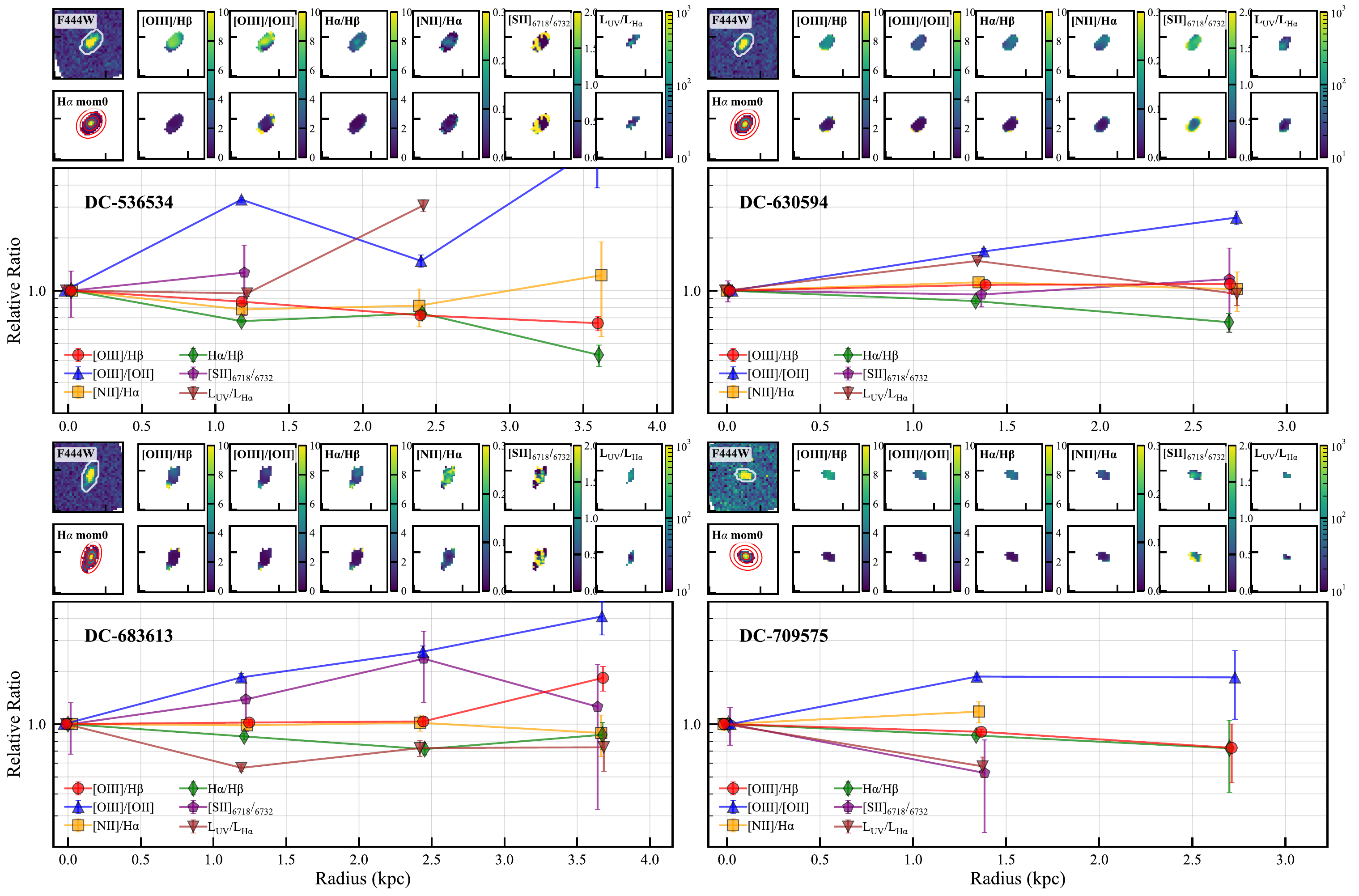}
\end{center}
\vspace{-0.2cm}
\caption{
Radial variations of emission-line ratios for our galaxies. 
The full compilation is provided in the Appendix~\ref{sec:appendix_sample}. 
For each target, we show $3''\times3''$ cutouts of the NIRCam F444W continuum and the H$\alpha$ moment-0 map (top left), 
together with maps of key diagnostic ratios (first row) and ratio errors (second row) in the top panels, left to right. 
The bottom panel displays the corresponding radially averaged profiles with $1\sigma$ uncertainties. 
The red ellipses overlaid on the H$\alpha$ map indicate the apertures adopted for the radial averaging, 
derived by fitting an elliptical Gaussian to the H$\alpha$ distribution.
Note that the radial coordinate of each data point represents the inner radius of the elliptical annulus, which does not affect the derived gradients (Section~\ref{sec:radial_grad}).
\label{fig:radial_ratio}}
\end{figure*}

The top panel of Figure~\ref{fig:radial_ratio} presents the 2D cutouts of ratio maps among key emission lines and the rest-frame UV continuum observed in four of our galaxies.
We also show the F444W image, the mask applied during the pixel-by-pixel SED fitting, and corresponding error maps derived by propagating the 1$\sigma$ flux uncertainties of the ratio measurements. To account for non-axisymmetric galaxy morphologies, we fit a 2D elliptical Gaussian profile to the H$\alpha$ moment-0 map, performing the least-square optimization to determine the centroid, major/minor axes, and position angles. 
Based on the best-fit shapes, we define the shape of radial annuli, and we adopt a constant width of $0\farcs2$ along the major axis ($\approx$1.3~kpc at $z=5$). Radially averaged fluxes are computed using median values, and uncertainties are estimated as the standard error of the mean within each annulus. The same figures for the full galaxy sample are provided in Appendix~\ref{sec:appendix_sample}. 

In the bottom panel of Figure~\ref{fig:radial_ratio}, we present the radially averaged ratios introduced above. To ensure robust interpretations, we display ratios only up to the annulus containing at least five pixels and where the measurements maintain an S/N of $\geq3$ for each ratio. Apart from H$\alpha$/H$\beta$, the displayed radial ratios are corrected for dust attenuation using radial H$\alpha$/H$\beta$ profiles. 
For annuli where the H$\alpha$/H$\beta$ ratio does not reach S/N $=3$, we adopt the value from the next inner annulus that satisfies the S/N threshold to perform the dust correction. 
We find clear radial trends in certain ratios, such as positive and negative trends in \oiii/\oii\ and H$\alpha$/H$\beta$, respectively, in all four galaxies, while this is not always the case among all our sample. 

To quantify these radial variations, we fit a single power law to each ratio in each galaxy. 
The best-fit parameters are obtained through least-squares optimization, with uncertainties estimated via Monte Carlo resampling in which the line ratios are perturbed by their measurement errors and the fitting repeated 100 times. 
The resulting slopes are summarized in Table~\ref{tab:line_ratio}, along with the weighted average for each ratio. 
\begin{table*}[!htbp]
\centering 
\setlength{\tabcolsep}{2pt}
\caption{Summary of radial gradient of key ratios among optical lines and UV continuum}
\label{tab:line_ratio}
\vspace{-0.2cm}
\begin{tabular}{lccccccc}
\hline
Target & $\Delta$[OIII]/H$\beta$ & $\Delta$[OIII]/[OII] & $\Delta$[NII]/H$\alpha$ & $\Delta$[SII]6718/6732 & $\Delta$H$\alpha$/H$\beta$ & $\Delta L_{\rm UV}$/$L_{\rm H\alpha}$ & $\Delta\log(\mathrm{O/H})$ \\
\hline
DC-417567\_a & $-0.059_{-0.006}^{+0.005}$ & $-0.097_{-0.006}^{+0.005}$ & $0.055_{-0.020}^{+0.018}$ & $0.029_{-0.103}^{+0.136}$ & $-0.037_{-0.007}^{+0.007}$ & \nodata & $0.09_{-0.04}^{+0.04}$ \\
DC-417567\_b & $-0.056_{-0.007}^{+0.006}$ & $0.002_{-0.010}^{+0.008}$ & $0.166_{-0.032}^{+0.038}$ & $0.135_{-0.114}^{+0.105}$ & $-0.041_{-0.008}^{+0.008}$ & \nodata & $0.04_{-0.04}^{+0.04}$ \\
DC-494763 & $-0.044_{-0.008}^{+0.007}$ & $0.199_{-0.012}^{+0.012}$ & $0.092_{-0.016}^{+0.014}$ & $-0.059_{-0.066}^{+0.069}$ & $-0.131_{-0.008}^{+0.009}$ & $-0.088_{-0.018}^{+0.021}$ & $0.04_{-0.04}^{+0.04}$ \\
DC-519281 & $-0.137_{-0.010}^{+0.010}$ & $0.087_{-0.029}^{+0.029}$ & $0.057_{-0.025}^{+0.025}$ & $0.135_{-0.114}^{+0.105}$ & $-0.109_{-0.012}^{+0.011}$ & $-0.199_{-0.019}^{+0.017}$ & $0.11_{-0.07}^{+0.07}$ \\
DC-536534 & $-0.054_{-0.007}^{+0.006}$ & $0.194_{-0.015}^{+0.014}$ & $-0.057_{-0.023}^{+0.025}$ & $0.097_{-0.142}^{+0.116}$ & $-0.101_{-0.007}^{+0.006}$ & $0.152_{-0.017}^{+0.015}$ & $0.09_{-0.12}^{+0.11}$ \\
DC-630594 & $0.021_{-0.012}^{+0.013}$ & $0.158_{-0.014}^{+0.014}$ & $0.029_{-0.020}^{+0.020}$ & $-0.005_{-0.055}^{+0.056}$ & $-0.051_{-0.013}^{+0.012}$ & $0.110_{-0.009}^{+0.008}$ & $-0.06_{-0.05}^{+0.06}$ \\
DC-683613 & $0.021_{-0.010}^{+0.009}$ & $0.172_{-0.015}^{+0.014}$ & $-0.003_{-0.017}^{+0.017}$ & $0.092_{-0.071}^{+0.080}$ & $-0.049_{-0.010}^{+0.012}$ & $-0.124_{-0.013}^{+0.012}$ & $-0.07_{-0.04}^{+0.04}$ \\
DC-709575 & $-0.033_{-0.013}^{+0.013}$ & $0.185_{-0.020}^{+0.020}$ & $0.059_{-0.051}^{+0.051}$ & $-0.207_{-0.188}^{+0.180}$ & $-0.047_{-0.015}^{+0.016}$ & \nodata & $0.01_{-0.07}^{+0.06}$ \\
DC-742174 & $-0.028_{-0.011}^{+0.011}$ & $0.015_{-0.018}^{+0.018}$ & $0.116_{-0.088}^{+0.086}$ & $-0.070_{-0.229}^{+0.238}$ & $-0.065_{-0.013}^{+0.013}$ & $0.051_{-0.010}^{+0.013}$ & $0.04_{-0.05}^{+0.04}$ \\
DC-842313\_a & $0.024_{-0.013}^{+0.012}$ & $0.493_{-0.029}^{+0.034}$ & $0.137_{-0.004}^{+0.004}$ & $0.024_{-0.020}^{+0.021}$ & $-0.167_{-0.012}^{+0.014}$ & $-0.046_{-0.026}^{+0.031}$ & $-0.04_{-0.05}^{+0.06}$ \\
DC-842313\_b & $-0.034_{-0.012}^{+0.013}$ & $-0.012_{-0.006}^{+0.006}$ & $0.052_{-0.018}^{+0.015}$ & $0.001_{-0.036}^{+0.032}$ & $-0.056_{-0.013}^{+0.013}$ & $0.063_{-0.010}^{+0.009}$ & $0.02_{-0.02}^{+0.02}$ \\
DC-848185\_a & $0.020_{-0.005}^{+0.005}$ & $0.085_{-0.009}^{+0.009}$ & $-0.030_{-0.012}^{+0.012}$ & $0.053_{-0.041}^{+0.047}$ & $-0.035_{-0.006}^{+0.006}$ & $0.088_{-0.016}^{+0.018}$ & $-0.06_{-0.05}^{+0.06}$ \\
DC-848185\_b & \nodata & \nodata & \nodata & \nodata & \nodata & \nodata & $-0.11_{-0.12}^{+0.12}$ \\
DC-873321\_a & $-0.107_{-0.007}^{+0.007}$ & $-0.142_{-0.013}^{+0.011}$ & $0.065_{-0.020}^{+0.018}$ & $0.094_{-0.075}^{+0.077}$ & $-0.051_{-0.007}^{+0.007}$ & $0.061_{-0.010}^{+0.010}$ & $0.07_{-0.04}^{+0.05}$ \\
DC-873321\_b & $-0.025_{-0.008}^{+0.008}$ & $-0.070_{-0.016}^{+0.014}$ & $0.095_{-0.038}^{+0.035}$ & $-0.093_{-0.093}^{+0.080}$ & $-0.002_{-0.009}^{+0.008}$ & $0.125_{-0.034}^{+0.031}$ & $0.05_{-0.03}^{+0.03}$ \\
DC-873756 & $-0.099_{-0.515}^{+0.446}$ & $-0.058_{-0.493}^{+0.519}$ & $-0.047_{-0.087}^{+0.077}$ & $0.116_{-0.489}^{+0.450}$ & $-0.156_{-0.141}^{+0.159}$ & $-0.436_{-0.083}^{+0.082}$ & \nodata \\
VC-5100541407\_a & $0.028_{-0.031}^{+0.030}$ & $0.071_{-0.022}^{+0.024}$ & $-0.008_{-0.033}^{+0.029}$ & $0.032_{-0.106}^{+0.117}$ & $-0.028_{-0.026}^{+0.030}$ & $-0.015_{-0.023}^{+0.021}$ & $-0.03_{-0.04}^{+0.04}$ \\
VC-5100541407\_b & \nodata & \nodata & \nodata & \nodata & \nodata & \nodata & \nodata \\
VC-5100822662\_a & $-0.077_{-0.013}^{+0.013}$ & $0.127_{-0.012}^{+0.012}$ & $0.047_{-0.020}^{+0.017}$ & $0.054_{-0.059}^{+0.059}$ & $-0.081_{-0.014}^{+0.014}$ & $0.140_{-0.028}^{+0.027}$ & $0.01_{-0.05}^{+0.05}$ \\
VC-5100822662\_b & \nodata & \nodata & \nodata & \nodata & \nodata & \nodata & $-0.05_{-0.15}^{+0.15}$ \\
VC-5100994794 & $-0.048_{-0.010}^{+0.009}$ & $-0.019_{-0.010}^{+0.009}$ & $-0.010_{-0.020}^{+0.018}$ & $-0.017_{-0.046}^{+0.043}$ & $-0.047_{-0.010}^{+0.011}$ & $0.140_{-0.006}^{+0.007}$ & $0.02_{-0.03}^{+0.02}$ \\
VC-5101218326 & $-0.043_{-0.043}^{+0.043}$ & $0.057_{-0.021}^{+0.027}$ & $-0.030_{-0.040}^{+0.029}$ & $0.003_{-0.129}^{+0.130}$ & $-0.111_{-0.056}^{+0.048}$ & $-0.198_{-0.016}^{+0.017}$ & $-0.05_{-0.05}^{+0.06}$ \\
VC-5101244930 & $-0.011_{-0.011}^{+0.010}$ & $-0.057_{-0.014}^{+0.013}$ & $-0.010_{-0.030}^{+0.029}$ & $-0.042_{-0.053}^{+0.059}$ & $-0.002_{-0.011}^{+0.011}$ & $-0.088_{-0.007}^{+0.007}$ & $0.03_{-0.04}^{+0.04}$ \\
VC-5110377875 & $-0.008_{-0.010}^{+0.010}$ & $-0.030_{-0.015}^{+0.015}$ & $0.032_{-0.017}^{+0.019}$ & $0.016_{-0.047}^{+0.052}$ & $-0.006_{-0.011}^{+0.011}$ & $0.000_{-0.009}^{+0.010}$ & $0.02_{-0.03}^{+0.03}$ \\ \hline
\textbf{Weighted avg.} & $-0.038 \pm 0.002$ & $0.013 \pm 0.003$ & $0.097 \pm 0.003$ & $0.017 \pm 0.012$ & $-0.057 \pm 0.002$ & $0.027 \pm 0.003$ & $0.020 \pm 0.009$ \\
\hline
\end{tabular}
\tablecomments{
The error of the weighted average is estimated by the standard error of the mean. All gradients are defined as 
$d(12+\log(\mathrm{O/H}))/dR$ and are expressed in units of dex\,kpc$^{-1}$.
The metallicity is measured by using the R3+O32 calibration (see Section~\ref{sec:zgas}) in this paper, while the impact of different metallicity calibrations will be presented in L.~Lee et al. (in prep.). 
}
\end{table*}
Although the gradients are generally shallow ($<0.1$~dex~kpc$^{-1}$), we find on average enhanced [O\,\textsc{iii}]/H$\beta$, [O\,\textsc{ii}]/[O\,\textsc{iii}], [S\,\textsc{ii}]$_{6732}$/[S\,\textsc{ii}]$_{6718}$, H$\alpha$/H$\beta$, and $L_{\rm H\alpha}/L_{\rm UV}$ toward the galaxy centers. 
These enhancements correspond to lower gas-phase metallicity \citep[e.g.,][]{curti2020, sanders2020, nakajima2022}, electron density, dust attenuation \citep[e.g.,][]{osterbrock1989}, and star-formation burstiness (SF$_{\rm burst}$; e.g., \citealt{faisst2019})\footnote{
This reflects differences in the luminosity-weighted timescales of star formation: $\sim$100~Myr for the rest-frame UV continuum versus $\sim$30~Myr for the H$\alpha$ emission.}, together with lower ionization parameter ($q_{\rm ion}$; e.g., \citealt{kewley2002}) in central regions, respectively. 
Such nearly flat yet systematic trends are consistent with recent \jwst\ spatially resolved spectroscopic studies of high-$z$ galaxies on kpc scales \citep[e.g.,][]{tripodi2024, venturi2024, ju2025}.
Although the coexistence of low $q_{\rm ion}$ and high SF$_{\rm burst}$ may appear counterintuitive, we note that [C\,\textsc{ii}] emission is strongly detected and peaks near the galaxy centers in most of our targets (Section~\ref{sec:cii_dust}). This implies that the central regions are gas-rich, potentially due to recent gas inflow, and such that even in the presence of high SF$_{\rm burst}$, the ionization parameter can be effectively reduced by dilution within the abundant gas. Taken together, these central properties, including the higher gas density implied by the low $q_{\rm ion}$, are consistent with an inside-out growth scenario for galaxies \citep[e.g.,][]{white1991}. 

\subsection{Metallicity gradient at $z=4-6$}
\label{sec:Zgrad}

Taking advantage of the spatially resolved emission-line ratios and their radial variations, we now turn to the radial gradients of $Z_{\rm gas}$, a key probe of galaxy growth and chemical mixing driven by the interplay among star formation, feedback, and gas accretion. In particular, in the classical inside-out growth scenario metallicity gradients are expected to be negative (i.e., higher at the center, decreasing outward) as central regions form and enrich first \citep[e.g.,][]{pilkington2012}. However, processes such as radial mixing via mergers or feedback-driven flows tend to flatten these gradients over time \citep[e.g.,][]{rupke2010}, while inflows of pristine gas into the outskirts can even invert the gradient, producing positive (increasing) profiles \citep[e.g.,][]{ceverino2016}. Thus, radial metallicity gradients serve as a sensitive diagnostic of inside-out growth, mixing, and gas flows. Note that here we provide an initial overview of the global trends in our sample, while a more detailed analysis of individual galaxies and their physical properties (e.g., stellar mass, sSFR, kinematics) and inspections on the effects of different line diagnostics will be presented in L.~Lee et al. (in prep.). Our metallicity gradient measurements are confirmed to be consistent with those of L.~Lee et al.\ generally within $\sim1\sigma$, despite the use of different metallicity calibrations and analysis methodologies.   

\begin{figure*}[!htbp]
\begin{center}
\includegraphics[trim=0cm 0cm 0cm 0cm, clip, angle=0,width=1.\textwidth]{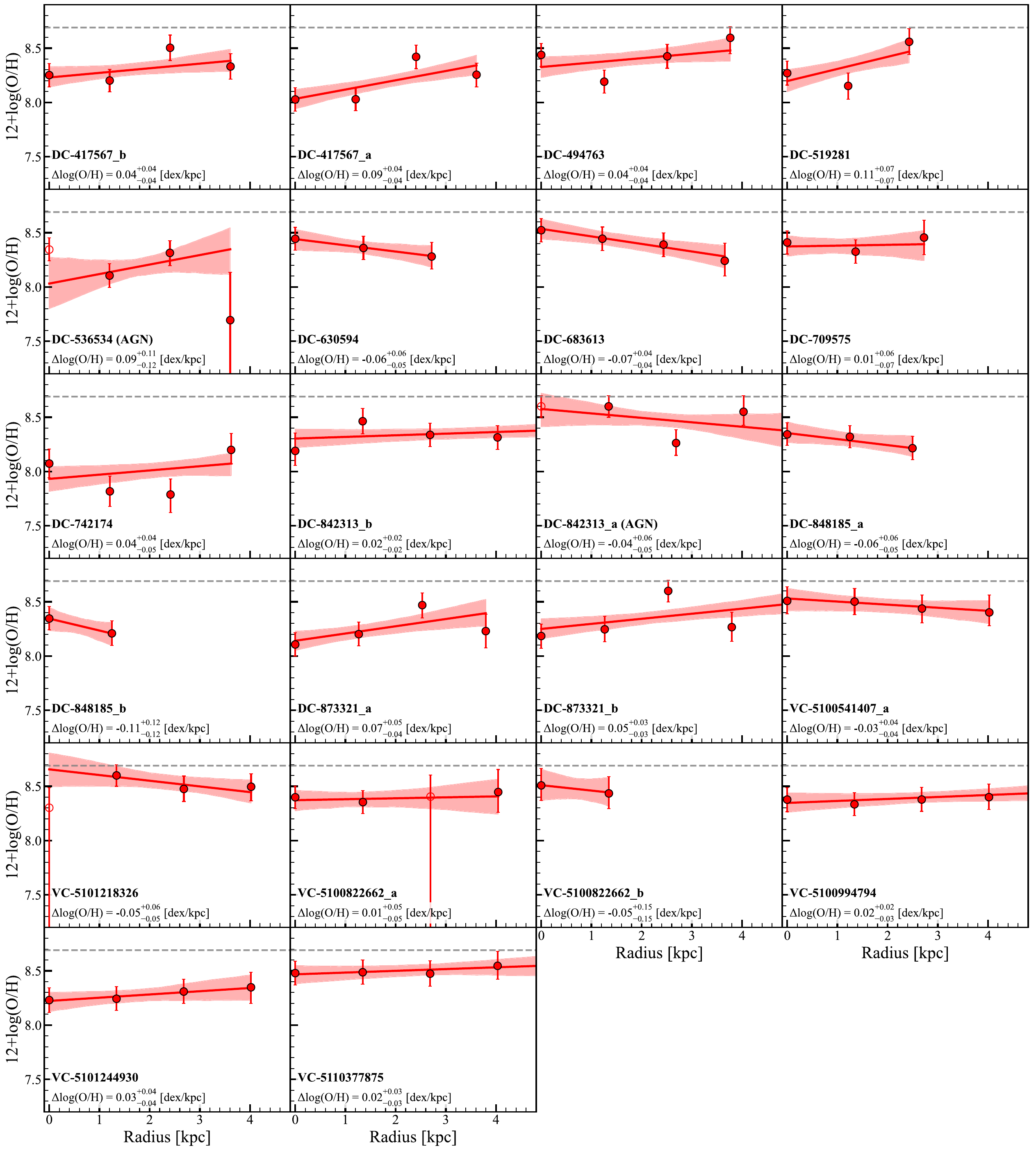}
\end{center}
\vspace{-0.2cm}
 \caption{ 
Metallicity gradients for individual galaxies based on the R3+O32 calibration. 
Each panel shows the radial profile of $12+\log({\rm O/H})$ with the best-fit slope (red line) 
and its $1\sigma$ confidence interval (shaded region), derived from Monte Carlo realizations. 
Galaxies with multiple spatial components within the IFU FoV are separated into each component (``a'' and ``b'') and fitted independently. 
The open red circles indicate data points with low S/N ($<3$) or central one of the AGN systems (DC-536534, DC-842313) that are not included in the fitting.  
The sample exhibits a broad range of behaviors, from negative to flat or even positive gradients, 
illustrating the diversity of chemical distribution patterns in star-forming galaxies at $z=4$–6.
\label{fig:zgrad_ind}}
\end{figure*}

\begin{figure*}[!htbp]
\begin{center}
\includegraphics[trim=0cm 0cm 0cm 0cm, clip, angle=0,width=0.9\textwidth]{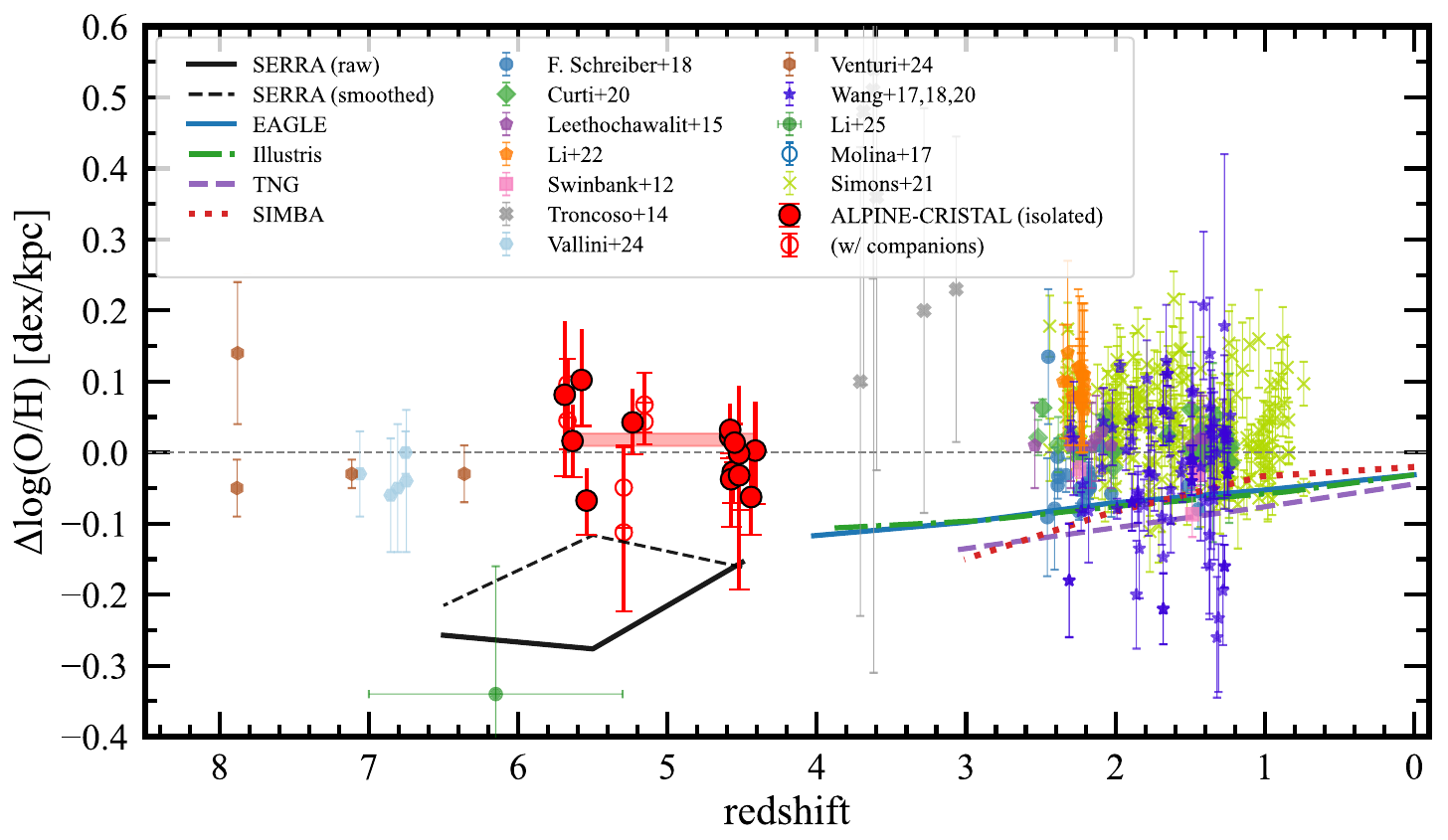}
\end{center}
\vspace{-0.2cm}
 \caption{
Gas-phase metallicity gradients ($\Delta \log({\rm O/H})/\Delta r$) as a function of redshift. 
Our ALPINE--CRISTAL measurements are shown in red circles, with filled symbols for isolated galaxies and open symbols for systems with companions. 
The dashed grey line marks the zero level for reference.
The red shaded region indicates the median and standard error of the mean for the full sample. 
Black solid and dashed curves show predictions from the SERRA cosmological zoom-in simulations \citep{pallottini2022, kohandel2024} at raw resolution and convolved to match the observational resolution, respectively. 
The comparison highlights that beam smearing flattens gradients by $\sim$0.1~dex, but the discrepancy with our observed values persists, pointing to efficient metal mixing in ALPINE--CRISTAL galaxies. 
Colored points represent literature measurements at lower and higher redshifts, while colored curves denote predictions from other cosmological simulations as compiled by \citet{garcia2025}. 
For consistency with our ALPINE--CRISTAL sample, we plot the simulation results for galaxies with 
$M_\star \simeq 10^{10}$--$10^{11}\,M_\odot$, corresponding to the stellar mass range of our observed galaxies.
Our results fill the observational gap at $z=4$--6 and support a scenario of little cosmic evolution in metallicity gradients.
\label{fig:Zgrad}}
\end{figure*}

Using the radial line ratios of [O\,\textsc{iii}]/H$\beta$ and [O\,\textsc{iii}]/[O\,\textsc{ii}], we compute $Z_{\rm gas}$ in each annulus in the same manner as Section \ref{sec:zgas} and derive metallicity gradients by applying the same linear fitting procedure described in Section~\ref{sec:radial_grad}. 
Note that we confirm the general consistency within 1$\sigma$ with another gradient measurement scheme by fitting to the pixel-by-pixel based values employed in L.~Lee et al. (in prep.).  
In Figure~\ref{fig:zgrad_ind}, we summarize the individual best-fit results, and Figure~\ref{fig:Zgrad} shows our metallicity gradient measurements as a function of redshift (red circles). 

Following the classification described in Section~\ref{sec:radial_grad}, isolated galaxies are shown as filled symbols, while galaxies with companions (labeled ``\_a'' and ``\_b'') are shown as open symbols. 
For comparison, we include recent measurements at both lower and higher redshifts (other colored symbols), as well as predictions from cosmological simulations (colored curves). 

First, we estimate the overall average value and examine potential differences within our sample. 
The weighted-average metallicity gradients for isolated galaxies, systems with neighbors, and the full sample are $0.01 \pm 0.01$, $0.03 \pm 0.01$, and $0.02 \pm 0.01$~dex~kpc$^{-1}$, respectively. 
Although the differences are modest, the trend suggests that galaxies with companions may reside in more massive dark-matter halos and may have recently accreted pristine gas, leading to slightly more positive metallicity gradients. 
This possible positive trend is independently confirmed by other measurements based on different metallicity calibrations (L.~Lee et al.\ in prep.; F.~Lopez et al.\ in prep.).

Second, we compare our results with previous observations. 
Before \jwst, intensive observational efforts have been achieved out to cosmic noon (i.e., $z\simeq$1--3), showing both negative, positive, and flat metallicity gradient results \citep{cresci2010, queyrel2012, swinbank2012b, troncoso2014, jones2013, jones2015, leethochawalit2016, wuyts2016, wang2017, wang2022, carton2018, schreiber2018b, curti2020}.  
Based on an analytical model approach using the FIR emission line properties, \cite{vallini2024} also report a flat gradient among star-forming galaxies at $z\simeq7$. 
Recent \jwst\ NIRSpec observations have enabled our capability of detecting the rest-frame optical emission lines out to $z\simeq$4--10, showing similarly flat \citep{venturi2024}, or tentative negative gradients depending on the choice of the metallicity calibration \citep{tripodi2024}. 
Nevertheless, these results so far are still consistent within little or no metallicity gradients across the cosmic times. 
An exception is \citet{li2025b}, where significantly negative metallicity gradients at $z>5$ are reported based on stacked NIRCam slitless spectra from ASPIRE \citep{fwang2023} and FRESCO \citep{oesch2023}. One possible reason is that their sample is naturally biased toward galaxies with strong [O\,\textsc{iii}]+H$\beta$ line emitters, i.e., bursty low-mass systems with $M_\star \lesssim 10^{9}\,M_\odot$, whereas our ALPINE--CRISTAL targets represent more massive main-sequence galaxies ($10^{9.5}$--$10^{10.5}\,M_\odot$; see A.~Faisst et al. submitted). 
Such selection differences likely contribute to the apparent tension due to the possible mass dependency for the metal mixing mechanisms, and indeed \citet{li2025b} find nearly flat gradients in their most massive bin ($M_\star \sim 10^{9}\,M_\odot$), consistent with our results.
Thus, our results are consistent within the range of these previous measurements, with flat or tentatively positive gradients and thus adding more weight on the little cosmic evolution of the metallicity gradients, at least among galaxies with $M_{\star}\gtrsim10^{9}\,M_{\odot}$. 

Third, we also compare our results with predictions from recent cosmological simulations of EAGLE \citep[e.g.,][]{tissera2019,tissera2022}, Illustris \citep[e.g.,][]{vogelsberger2014}, TNG \citep[e.g.,][]{hemler2021}, SIMBA \citep[e.g.,][]{dave2019}, and SERRA \citep[e.g.,][]{pallottini2022,kohandel2023}. 
Among them, we highlight the SERRA simulations \citep[e.g.,][]{pallottini2022,kohandel2023}, which provide mock galaxies at the same redshift range as our ALPINE--CRISTAL sample. 
To ensure a fair comparison, we randomly select 130 SERRA galaxies with $M_{\star}\simeq10^{10}-10^{11}\,M_{\odot}$ at $z=4$--7 and process their mock maps following the same procedures as applied to our observed data. 
Specifically, we convolve the intrinsic $M_{\star}$ maps with a PSF to mimic the spatial resolution of NIRSpec IFU, add noise corresponding to the typical peak S/N of our maps, and generate segmentation masks using {\tt SExtractor}. 
We then determine galaxy geometry (centroid, axis ratio, and position angle) from the $M_\star$ maps using two-dimensional elliptical Gaussian fitting, derive radial profiles from the masked metallicity maps (with and without PSF smoothing), and measure gradients by fitting a weighted linear relation. 
This procedure allows us to characterize the intrinsic gradients in SERRA as well as the impact of observational systematics such as beam smearing in a manner directly comparable to our data. 

We find that SERRA galaxies are generally characterized by clear negative gradients of $\sim-0.3$ to $-0.2$~dex~kpc$^{-1}$. 
Beam smearing flattens these gradients by $\sim$0.1~dex~kpc$^{-1}$ (see black dashed vs.\ solid curves in Figure~\ref{fig:Zgrad}), as expected, but this is still insufficient to reproduce the nearly flat ($\sim$0) gradients observed in the ALPINE--CRISTAL galaxies (see also the independent test in L.~Lee et al., in prep.). 
At lower redshifts, other cosmological simulations likewise predict negative gradients for galaxies of similar stellar mass. 
The gradients, however, depend sensitively on how metal-mixing processes from inflows, outflows, and mergers are implemented; the discrepancy with our data may therefore indicate that efficient metal mixing taking place in real galaxies is not yet fully captured in current simulations.
Nevertheless, we emphasize that the metallicities in simulated galaxies are based on intrinsic O/H values, and a direct comparison with observational metallicities derived from strong-line calibrations is still lacking. 
A self-consistent treatment of metallicity derivations, together with a systematic investigation of the relevant physical mechanisms across cosmic time, will be essential to fully understand the evolution of metallicity gradients and their physical origin.

Overall, these consistent results of the nearly flat gradient suggest that the vigorous metal mixing processes already in place by $z>6$ continue at $z=4$--6, with little evidence for any substantial change in their efficiency toward cosmic noon. This consistent trend implies that inflows, outflows, and interactions have regulated the internal chemical structures of galaxies over more than two billion years of cosmic time, leaving metallicity gradients consistently flat or only mildly negative across epochs. Importantly, the redshift range $z\simeq4$--6 has so far been a missing link in the study of metallicity gradients. Our measurements provide the first statistical reference sample at these epochs, bridging our understanding of chemical enrichment from the local universe through cosmic noon to the era of reionization.

\subsection{Resolved Fundamental Relation at $z=4$--6}
\label{sec:rFMR}

The fundamental metallicity relation (FMR) connects stellar mass ($M_\star$), star formation rate (SFR), and gas-phase metallicity ($Z_{\rm gas}$) in galaxies, capturing the interplay between gas accretion, star formation, and chemical enrichment. In the local Universe, this relation has been well studied both in integrated and spatially resolved forms (e.g., \citealt{mannucci2010, belfiore2017, baker2023}), and its resolved version---the resolved FMR (rFMR)---has proven useful in understanding local ISM physics and feedback cycles on kiloparsec scales. Studying whether such a relation exists at earlier cosmic times provides a window into the regulation of star formation in young galaxies and the prevalence of gas inflows and outflows during the early stages of galaxy evolution, together with the metallicity gradient measurements (Section~\ref{sec:Zgrad}). 

To examine the rFMR at $z = 4$--6, we derive spatially resolved maps of $M_\star$ and SFR via pixel-by-pixel SED fitting (Section~\ref{sec:sed}), while gas-phase metallicities are measured from emission line flux ratios (Section~\ref{sec:zgas}). The SED fitting incorporates JWST/NIRCam photometry and NIRSpec emission line maps (H$\alpha$, H$\beta$, [O\,\textsc{iii}]), ensuring a self-consistent estimate of stellar and nebular properties on a spaxel-by-spaxel basis. Because a few galaxies in our sample are reported to be broad-line AGN candidates, we mask out central pixels showing obvious broad-line contamination (typically 2--3 pixels across the core) to minimize the impact on metallicity measurements and preserve the validity of rFMR comparisons with purely star-forming regions (see Section~\ref{sec:zgas}).

\begin{figure*}[!htbp]
\begin{center}
\includegraphics[trim=0cm 0cm 0cm 0cm, clip, angle=0,width=1.\textwidth]{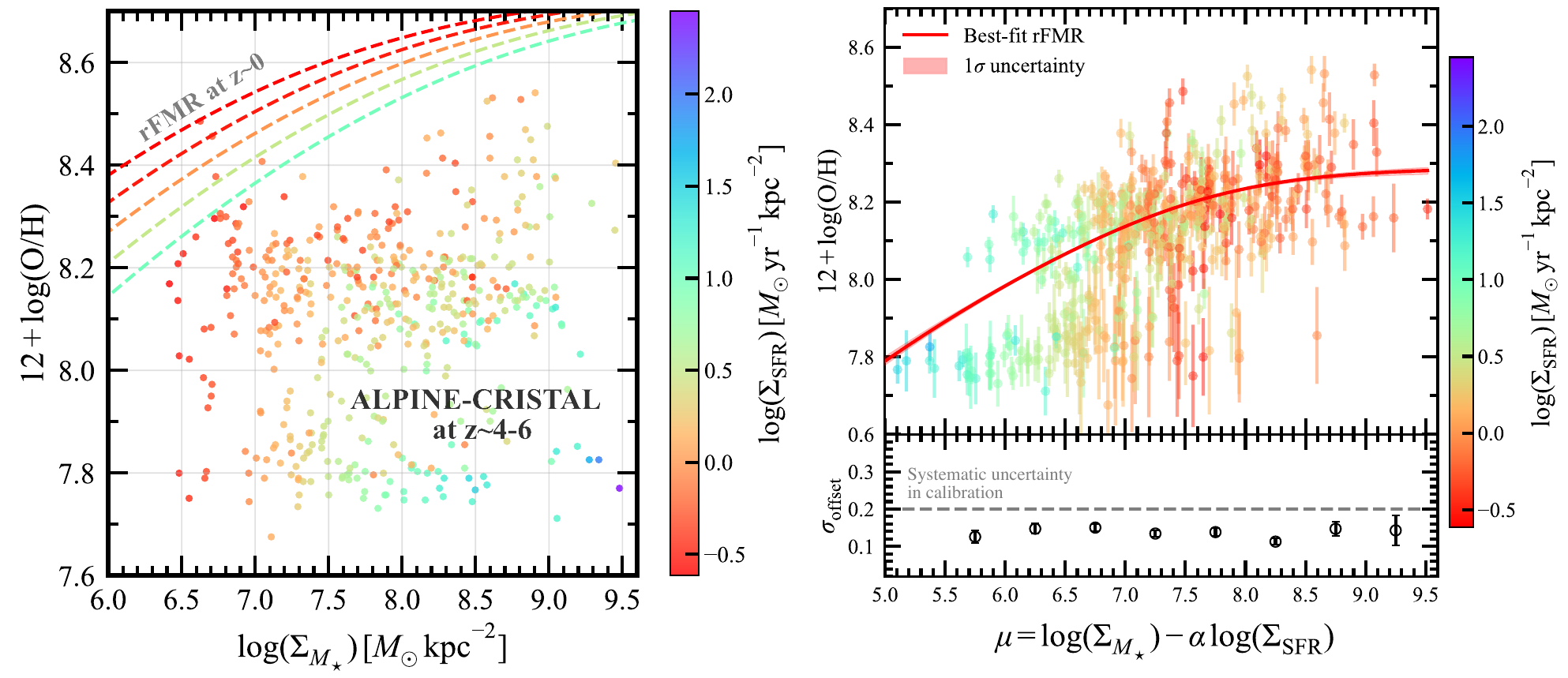}
\end{center}
\vspace{-0.2cm}
 \caption{
\textbf{\textit{Left}}:
Resolved Fundamental Metallicity Relation (rFMR) at $z = 4$--6, showing gas-phase metallicity as a function of stellar mass surface density and star formation rate surface density. 
Each point corresponds to a spatial pixel (spaxel) from our sample galaxies with sufficient S/N. 
The dashed lines indicate the best-fit rFMR relation in local galaxies \citep{baker2023}. 
\textbf{\textit{Right}}:
Best-fit 3D surface of the rFMR at $z=4$--6, based on the projection parameter $\alpha=2.1$. 
The solid line indicates the best-fit rFMR relation, and the shaded region represents the $1\sigma$ uncertainty. 
The bottom panel denotes residuals of individual spaxels from the best-fit plane. 
The tight correlation, with scatter comparable to or smaller than the typical systematic uncertainty of strong-line metallicity calibrations ($\simeq$0.2~dex; e.g., \citealt{li2025b}), supports the presence of a resolved FMR at early cosmic epochs. A horizontal branch is also observed at low metallicity and can be interpreted as a signature of recent pristine gas inflows that dilute the ISM metallicity.
\label{fig:rfmr}}
\end{figure*}

The left panel of Figure~\ref{fig:rfmr} presents the resulting distribution of spatially resolved metallicity as a function of the stellar mass surface density ($\Sigma_\star$), color coded by the SFR surface density ($\Sigma_{\rm SFR}$).
For comparison, we overlay the best-fit plane for the local rFMR derived in \citet{baker2023}, extrapolated to the dynamic range of our high-$z$ data.

We find that, at fixed $\Sigma_\star$, our sample exhibits systematically lower metallicities compared to the local rFMR plane, with significant dispersions. This offset is consistent with the known redshift evolution of the mass-metallicity relation \citep[e.g.,][]{maiolino2008, sanders2020, papovich2022a, nakajima2023, curti2024}, and reflects the less chemically enriched state of galaxies at earlier cosmic times. However, more importantly, we also find a trend consistent with the local rFMR: at fixed $\Sigma_\star$, spaxels with higher $\Sigma_{\rm SFR}$ tend to exhibit lower metallicities. 

This inverse dependence of $Z_{\rm gas}$ on $\Sigma_{\rm SFR}$ has traditionally been interpreted as a signature of recent inflows of low-metallicity gas, which simultaneously dilute the existing ISM metals and fuel bursts of star formation (e.g., \citealt{cresci2010, dave2012, forbes2014}). However, recent studies have shown that the SFR–metallicity correlation is not solely explained by gas content or inflow-driven dilution. \citet{baker2023} demonstrate that in local galaxies, the SFR (especially global SFR) has a direct role in modulating the gas-phase metallicity, potentially via metal-rich galactic outflows powered by stellar feedback. In this framework, intense star formation episodes not only reflect recent gas accretion but also actively remove metals through powerful winds, lowering the observed metallicity in high-$\Sigma_{\rm SFR}$ regions \citep[e.g.,][]{hamel-bravo2024}.

Our results suggest that both mechanisms, metal-poor gas inflows and SFR-driven outflows, may already be taking place at $z\sim4$--6. The presence of an rFMR at this epoch, with a similar inverse $Z$–$\Sigma_{\rm SFR}$ correlation as observed locally, indicates that the same regulatory processes shaping the ISM at late times were already in place in early galaxies. Over time, these galaxies are likely to increase their metal content as the inflow rate declines and chemical enrichment proceeds, eventually approaching the more chemically mature state seen at lower redshift.

To further quantify the rFMR and compare it to studies in the local Universe, we adopt a projected parameter $\mu$, defined as a linear combination of the stellar mass and SFR surface densities:
\begin{equation}
    \mu = \log \Sigma_\star - \alpha \log \Sigma_{\rm SFR},
\end{equation}
where $\alpha$ is a free parameter to be optimized. This parameterization effectively captures the curvature of the rFMR surface and allows a more compact representation of the data.

Following the analytic formulation introduced by \citet{curti2020b} and used in \citet{baker2023}, we model the rFMR using $\mu$ with the functional form:
\begin{equation}
    12 + \log({\rm O/H}) = \rho - \frac{\zeta}{\psi} \log_{10} \left[1 + \left( \frac{\mu}{\omega} \right)^{-\psi} \right],
\end{equation}
where $\rho$, $\zeta$, $\psi$, and $\omega$ control the normalization, low-$\mu$ slope, transition sharpness, and turnover point, respectively.

We fit this model to our resolved measurements by minimizing the residuals between the observed metallicities and those predicted by the relation, treating all five parameters ($\rho$, $\zeta$, $\psi$, $\omega$, $\alpha$) as free. The best-fit values and their $1\sigma$ uncertainties are: $\rho = 8.28 \pm 0.06$, $\zeta = 2.5 \pm 0.7$, $\psi = 13 \pm 11$, $\omega = 8.0 \pm 0.4$, and $\alpha = 2.1 \pm 0.2$. 

The right panel of Figure~\ref{fig:rfmr} shows the best-fit fundamental plane (top) and the average residual scatter in each $\mu$ bin (bottom). 
The residuals are well within the typical systematic uncertainty of strong-line metallicity calibrations ($\simeq$0.2~dex; e.g., \citealt{li2025b}). 
While the intrinsic dispersion may physically increase due to the growing burstiness, merger activity, and shallower dark-matter potentials of galaxies toward higher redshifts \citep[e.g.,][]{pallottini2025}, 
our results indicate that the observed scatter can be accounted for by calibration systematics, and that our spaxel-level data points lie in good agreement with the inferred $\mu$--metallicity relation. 
As also seen in the left panel, this does not imply an exact match to the local FMR, but it does show that galaxies at $z\sim4$--6 already follow a similar plane linking stellar mass, SFR, and metallicity. 

Notably, the best-fit projection parameter of \(\alpha = 2.1 \pm 0.2\) is substantially larger than the values typically reported at lower redshift ($\alpha\simeq0.3$--0.6; e.g., \citealt{mannucci2010,curti2020b,baker2023}). 
Such a high $\alpha$ implies that, in our $z\sim4$--6 sample, metallicity depends much more strongly on $\Sigma_{\rm SFR}$ than on $\Sigma_\star$. 
This trend is qualitatively consistent with the findings of \citet{bulichi2023}, who reported that in local dwarf galaxies the anti-correlation between $\Sigma_{\rm SFR}$ and metallicity is particularly tight. 
Therefore, our results suggest that in environments characterized by low metallicity and high specific star formation, local metallicity is more strongly regulated by short-timescale processes -- dilution by inflows of pristine gas or removal of metals via SFR-driven outflows -- prior to the long-term build-up of stellar mass becoming significant. 
This is also in line with the nearly flat metallicity gradients we observe (Section~\ref{sec:Zgrad}), which is reflected by efficient mixing of metals throughout the ISM.  

We note that the elevated gas turbulence observed in high-$z$ galaxies is expected to further enhance the efficiency of metal mixing, contributing to the regulation of local metallicity \citep[e.g.,][]{jones2013,wang2017,wisnioski2015,lee2025}. 
Such turbulent mixing has been discussed in several studies at $z\sim1$--2 and represents another important mechanism, in addition to inflows and outflows, that shapes the internal chemical structure of galaxies. 

Interestingly, we also detect a horizontal branch at the low-metallicity regime (12+$\log$(O/H)$\lesssim8.0$). 
This feature can be explained by recent pristine gas inflows that dilute the gas-phase metallicity in galaxies. 
Such tight connections between the emergence of the horizontal branch and the evidence of recent pristine gas inflows have been confirmed in local metal-poor galaxies through intensive IFU observations \citep{nakajima2024}. 
Thus, the presence of the horizontal branch supports the picture that active metal mixing already takes place at least through gas inflows in our galaxies at $z=$4--6. 

These results suggest that the rFMR observed in our $z=4$--6 galaxies reflects the same fundamental mechanisms that operate in local low-mass systems, highlighting a continuity of underlying chemical regulation processes across cosmic time. 
Our elevated $\alpha$ value and the presence of the horizontal branch do not merely reflect extreme conditions at $z\sim4$--6, but instead represents an early phase of galaxy formation governed by processes that are likewise observed in metal-poor dwarf galaxies in the local Universe today.

\section{Summary}
\label{sec:summary}

We present \jwst/NIRSpec G235M and G395M (or G395H) IFU observations of 18 main-sequence galaxies at $z=4$--6 from the ALPINE--CRISTAL survey, providing the first statistical view of spatially resolved chemical distributions at these epochs. Our optimized data reduction, including background subtraction, light-leakage masking, stripe removal, error rescaling, and astrometric refinement. Strong rest-frame optical emission lines (e.g., H$\alpha$, H$\beta$, \oiii, \oii) are robustly detected in nearly all galaxies, enabling us to achieve reliable emission-line mapping on kiloparsec scales. The main results are summarized as follows:

\begin{itemize}
    \item The 2D maps of metallicity and dust attenuation reveal substantial diversity: some galaxies show centrally enhanced values coincident with stellar mass peaks, while others exhibit flat or off-centered distributions. The [C\,\textsc{ii}] and dust continuum generally trace stellar mass peaks, though with notable exceptions, pointing to complex ISM conditions.
    \item Radial trends of emission-line ratios indicate mild central enhancements in [O\,\textsc{iii}]/H$\beta$, [O\,\textsc{ii}]/[O\,\textsc{iii}], [S\,\textsc{ii}]$_{6732}$/[S\,\textsc{ii}]$_{6718}$, H$\alpha$/H$\beta$, and $L_{\rm H\alpha}/L_{\rm UV}$, consistent with higher electron density, dust obscuration, and bursty star formation together with lower  gas-phase metallicity and ionization parameters in central regions. These signatures support an inside-out growth scenario moderated by recent inflows of pristine gas.
    \item Despite the mild trend, the median metallicity gradient remains nearly flat, $\Delta\log({\rm O/H}) = 0.02 \pm 0.01$ dex kpc$^{-1}$. This flatness implies efficient mixing of metals via inflows, outflows, and mergers, and is consistent with the little or no cosmic evolution in gradients observed from $z \sim 0$ to $z \sim 6$.
    \item We establish the resolved Fundamental Metallicity Relation (rFMR) at $z=4$--6. 
    Metallicity is well described by a plane in stellar mass and SFR surface densities, with a stronger dependence on $\Sigma_{\rm SFR}$ (projection parameter $\alpha \simeq 2.1$) than in local galaxies. 
    This elevated $\alpha$ indicates the enhanced role of short-timescale processes---inflows of metal-poor gas and SFR-driven outflows---in regulating metallicity under low-$Z$, high-sSFR conditions. 
    We also identify a low-metallicity horizontal branch in the rFMR, similar to that seen in local metal-poor galaxies, which likely traces recent pristine gas inflow.
    \item These results demonstrate that the regulatory cycle linking gas accretion, feedback, and star formation was already vigorous $\sim1$~Gyr after the Big Bang. The nearly flat metallicity gradients and the presence of an rFMR reveal a continuity of the governing physical mechanisms from the era of reionization through cosmic noon to the present day, with star-formation-driven processes playing an increasingly prominent role under the conditions of the early universe.
\end{itemize}

We thank Micaela Bagley and Susan Kassin for valuable advice on data reduction, 
Giacomo Venturi and Mirko Curti for kindly sharing their compilation of literature data, 
and Roberto Maiolino and Ryan Sanders for insightful discussions on metallicity calibrations and spatially-resolved fundamental relation. 
This work is based on observations and archival data made with the {\it Spitzer Space Telescope}, which is operated by the Jet Propulsion
Laboratory, California Institute of Technology, under a contract with NASA along with archival data from the NASA/ESA 
{\it Hubble Space Telescope}. 
The Dunlap institute is funded through an endowment established by the David Dunlap family and the University of Toronto. 
H\"U acknowledges funding by the European Union (ERC APEX, 101164796). Views and opinions expressed are however those of the authors only and do not necessarily reflect those of the European Union or the European Research Council Executive Agency. Neither the European Union nor the granting authority can be held responsible for them.
MB gratefully acknowledges support from the ANID BASAL project FB210003. 
This work was supported by the French government through the France 2030 investment plan managed by the National Research Agency (ANR), as part of the Initiative of Excellence of Université Côte d'Azur under reference number ANR-15-IDEX-01.
We acknowledge the Lorentz Center for giving us the opportunity and infrastructure to organize a successful and very stimulating workshop, which discussions helped to realize this work.
MA is supported by FONDECYT grant number 1252054, and gratefully acknowledges support from ANID Basal Project FB210003 and ANID MILENIO NCN2024\_112
RJA was supported by FONDECYT grant number 1231718 and by the ANID BASAL project FB210003. 
\tcb{JS is supported by JSPS KAKENHI (JP22H01262). }

The JWST and HST data presented in this article were obtained from the Mikulski Archive for Space Telescopes (MAST) at the Space Telescope Science Institute.
These observations are associated with programs, HST-GO-13641 (\dataset[doi: 10.17909/xne1-7v26]{https://doi.org/10.17909/xne1-7v26}), JWST-GO-01727 (\dataset[doi: 10.17909/ph8h-qf05]{https://doi.org/10.17909/ph8h-qf05}), JWST-GO-03045 (\dataset[doi: 10.17909/cqds-qc81]{https://doi.org/10.17909/cqds-qc81}), and JWST-GO-04265 (\dataset[doi: 10.17909/wac6-9741]{https://doi.org/10.17909/wac6-9741}).
The reduced \jwst/NIRCam data will be available via \url{http://alpine.ipac.caltech.edu/}.

\software{\texttt{Astropy} \citep{astropy2013,astropy2018,astropy2022},
          {\sc Source Extractor} \citep{bertin1996}, 
          \texttt{Prospector} \citep{leja2017, johnson2021}
          }

\appendix

\section{Additional Masking of light leakage}
\label{sec:appendix_mask}

During inspection of the Stage~2 products, we identified non-negligible light leakage in the detector images that was not fully masked by the default pipeline. These residual features stem from both known failed open shutters and intermittent ones, and were not completely removed by the pipeline masking based on the known stacked shutter information.
To address this, we visually inspected all dithered exposures and manually added masks to eliminate the remaining light leakage. 

\begin{figure*}
\begin{center}
\includegraphics[trim=0cm 0cm 0cm 0cm, clip, angle=0,width=1.\textwidth]{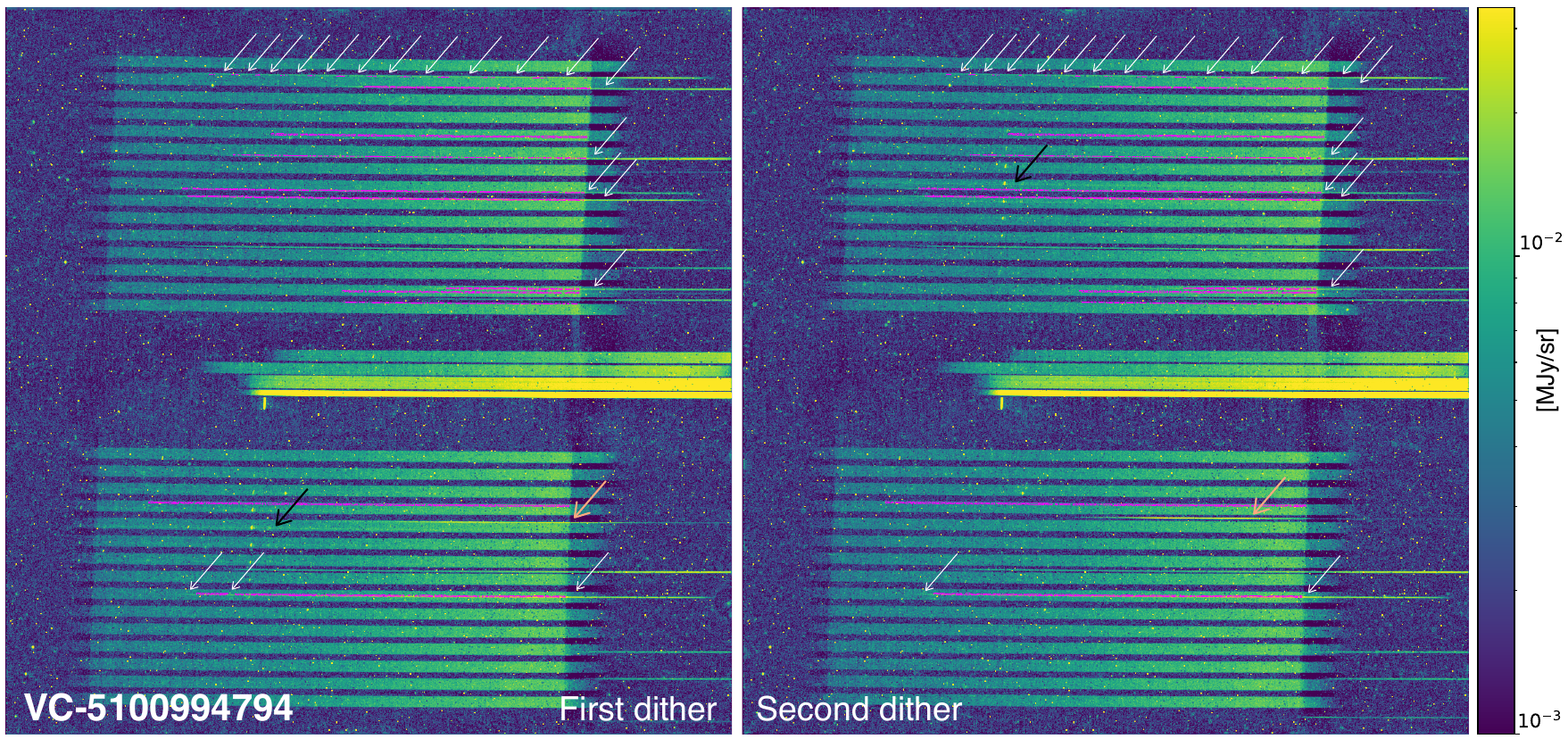}
\end{center}
\vspace{-0.2cm}
\caption{
NIRSpec/IFU detector images for the two dither positions in visit VC-5100994794, using a 2-point sparse cycling dither pattern. Each stripe corresponds to dispersed spectra from a row of spaxels. The magenta contours indicate pixels masked by the default pipeline based on known failed open shutters. White and orange arrows mark residual light leakage from known and intermittent failed shutters, respectively, while black arrows indicate continuum from actual sources, whose positions shift between dithers. The identical wavelength coverage in both dithers, combined with slight positional offsets that improves spatial sampling and helps mitigating the systematics. 
\label{fig:manual_mask}
}
\end{figure*}

Figure~\ref{fig:manual_mask} shows an example from the two dither positions in visit VC-5100994794. Pixels masked by the default pipeline (based on known failed open shutters) are shown as magenta contours. We also highlight three key features: (1) residual light leakage from known failed shutters that was not fully masked (white arrows), (2) light leakage from intermittent failed shutters (orange arrows), and (3) actual source continuum (black arrows).
We find that (1) occurs in most cases, whereas (2), particularly for those falling on dispersed spectrum regions, takes place only occasionally (roughly once every few frames).
Importantly, the true source continuum shifts in detector position between dithers, particularly in our 2-point sparse cycling pattern, whereas the light leakage remains the same positions. This allows us to distinguish between genuine astrophysical signal and instrument-related artifacts. We applied additional rectangular masks to the positions of the remaining light leakage (indicated by the white and orange arrows) prior to Stage~3 processing. 
We note that any light leakage that falls outside the IFU spectral traces, or outside the wavelength coverage of those traces, is naturally excluded in the final data cube building process, so we did not aggressively mask such regions.

\section{Astrometry and flux comparison using NIRCam and NIRSpec}
\label{sec:appendix_additional}

As described in Section~\ref{sec:post}, we applied additional steps beyond the pipeline reduction to verify the quality of our data, including tests on astrometry and absolute flux calibration. 

Table~\ref{tab:offset} summarizes the spatial offsets between the peaks in the NIRCam/F277W and F444W images and those in the corresponding pseudo-F277W and F444W images generated by collapsing the NIRSpec IFU cubes with the filter response curves. Based on these measured offsets, we corrected the astrometry of the NIRSpec IFU so that the peaks in the pseudo-continuum maps align with the NIRCam peak positions. 

\begin{figure}[!htbp]
\begin{center}
\includegraphics[trim=0cm 0cm 0cm 0cm, clip, angle=0,width=0.5\textwidth]{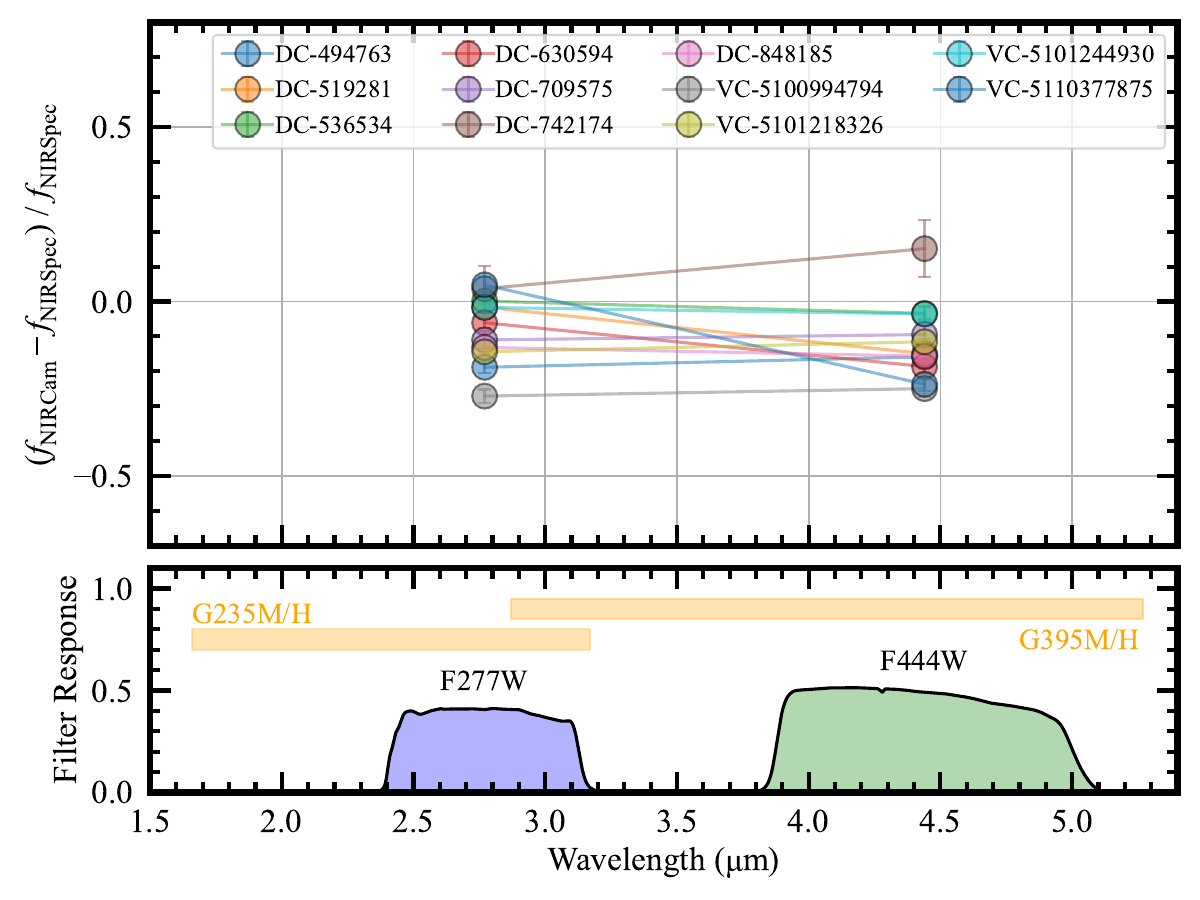}
\end{center}
\vspace{-0.2cm}
 \caption{
Flux comparison between NIRSpec and NIRCam by using the same aperture on the pseudo and actual NIRCam maps. 
To mitigate unknown impacts from the potential remaining astrometry offset, we use only visually isolated galaxies for this comparison. 
The NIRCam filter response and the grating range of the NIRSpec G235M and G395M are also shown in the bottom panel. 
We find that the flux is consistent within $\simeq$~20--30\% between both instruments. 
This dispersion likely stems from a combination of several uncertainties, such as the absolute flux calibrations, background subtractions, and astrometry corrections in both NIRCam and NIRSpec. 
\label{fig:flux_comp}}
\end{figure}

Figure~\ref{fig:flux_comp} presents a comparison between the NIRCam photometric measurements and the NIRSpec IFU fluxes extracted within the same apertures and convolved with the appropriate NIRCam filter response curves. We find that the typical flux differences are close to zero, although they can reach up to $\sim$30\%. The origin of these discrepancies is uncertain, and may involve absolute flux calibration, background subtraction, or remaining astrometric errors in either NIRSpec or NIRCam. Given the overall consistency between the two instruments, we do not propagate this potential systematic uncertainty from the absolute flux calibration into our subsequent analyses.

\begin{table}[!htbp]
\centering
\footnotesize
\setlength{\tabcolsep}{1.5pt}
\caption{Summary of our astrometry offset measurements}
\label{tab:offset}
\vspace{-0.2cm}
\begin{tabular}{lcccccc}
\hline
Target & \multicolumn{2}{c}{$\Delta$RA} & \multicolumn{2}{c}{$\Delta$Dec} & $\langle\Delta$RA$\rangle$ & $\langle\Delta$Dec$\rangle$ \\
 & G235M & G395M & G235M & G395M &   \multicolumn{2}{c}{G235M+G395M} \\
 & (arcsec) & (arcsec) & (arcsec) & (arcsec) & (arcsec) & (arcsec) \\
\hline
DC-417567 & 0.03 & 0.02 & 0.17 & 0.17 & 0.02 & 0.17 \\
DC-494763 & 0.09 & 0.07 & 0.18 & 0.19 & 0.08 & 0.19 \\
DC-519281 & 0.08 & 0.08 & 0.15 & 0.15 & 0.08 & 0.15 \\
DC-536534 & 0.10 & 0.07 & 0.20 & 0.18 & 0.09 & 0.19 \\
DC-630594 & 0.06 & 0.06 & 0.14 & 0.14 & 0.06 & 0.14 \\
DC-683613 & 0.03 & 0.04 & 0.17 & 0.15 & 0.04 & 0.16 \\
DC-709575 & 0.07 & 0.09 & 0.15 & 0.15 & 0.08 & 0.15 \\
DC-742174 & 0.08 & 0.06 & 0.13 & 0.15 & 0.07 & 0.14 \\
DC-842313 & 0.07 & 0.07 & 0.14 & 0.14 & 0.07 & 0.14 \\
DC-848185 & 0.06 & 0.12 & 0.17 & 0.15 & 0.09 & 0.16 \\
DC-873321 & 0.09 & 0.08 & 0.17 & 0.19 & 0.09 & 0.18 \\
DC-873756 & 0.08 & 0.04 & 0.24 & 0.23 & 0.06 & 0.23 \\
VC-5100541407 & 0.02 & 0.04 & 0.16 & 0.18 & 0.03 & 0.17 \\
VC-5100822662 & 0.08 & 0.08 & 0.16 & 0.18 & 0.08 & 0.17 \\
VC-5100994794 & 0.11 & 0.10 & 0.16 & 0.15 & 0.11 & 0.16 \\
VC-5101218326 & 0.10 & 0.10 & 0.15 & 0.14 & 0.10 & 0.15 \\
VC-5101244930 & 0.10 & 0.16 & 0.14 & 0.11 & 0.13 & 0.12 \\
VC-5110377875 & 0.08 & 0.10 & 0.14 & 0.16 & 0.09 & 0.15 \\
\hline
Average & 0.07 & 0.08 & 0.16 & 0.16 & 0.08 & 0.16 \\
\hline
\end{tabular}
\end{table}

\section{Compilation of 1D spectra, 2D moment-0, and radial line-ratio maps}
\label{sec:appendix_sample}
In this Appendix we provide the full compilation of the NIRSpec IFU moment-0 maps for the rest-optical emission lines together with the ALMA dust continuum and [C\,\textsc{ii}] 158\,$\mu$m maps (Figure~\ref{fig:mom0_comb}), and the radial emission-line ratio maps and profiles (Figure~\ref{fig:radial_ratio}) for our galaxy sample, 
complementing the examples shown in Figures~\ref{fig:comb_example} and \ref{fig:radial_ratios_appendix}. 

\begin{figure*}
    \centering
    \includegraphics[width=0.8\textwidth]{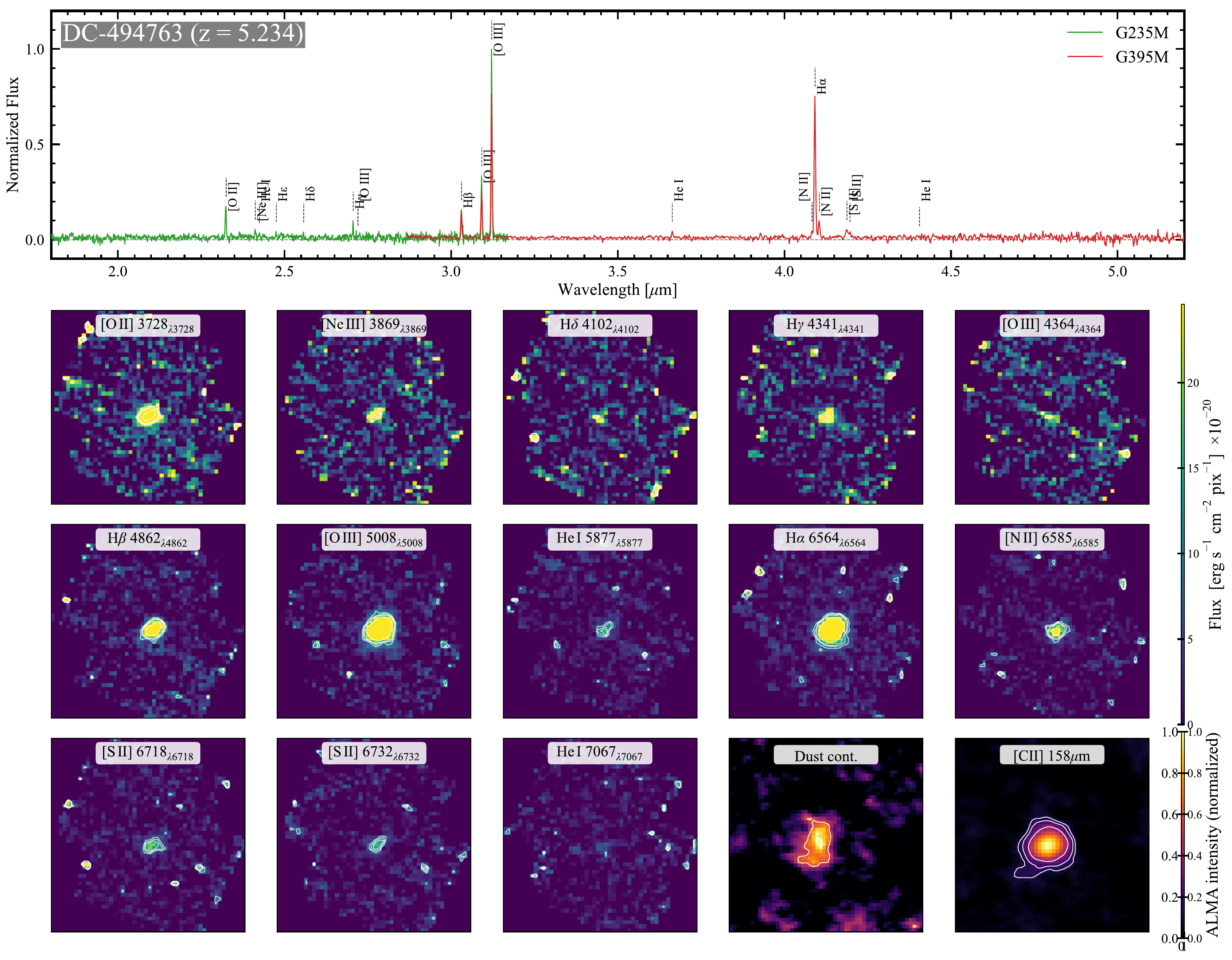}
    \caption{
    Same as Figure~\ref{fig:comb_example}, but showing all remaining targets in the sample. 
    }
    \label{fig:mom0_comb}
\end{figure*}

\addtocounter{figure}{-1}
\begin{figure*}
    \centering
    \includegraphics[width=0.8\textwidth]{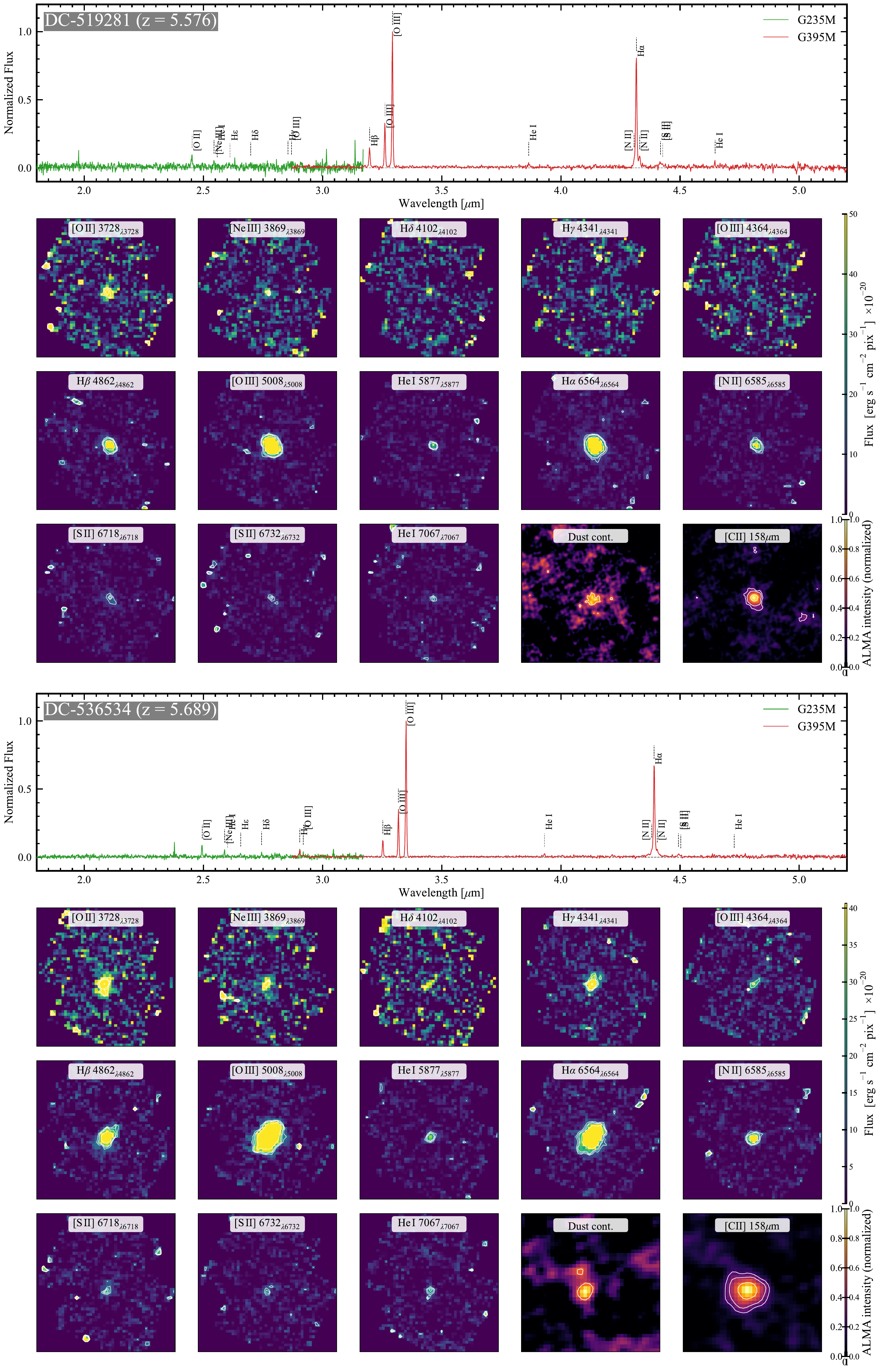}
    \caption{(continued)}
\end{figure*}

\addtocounter{figure}{-1}
\begin{figure*}
    \centering
    \includegraphics[width=0.8\textwidth]{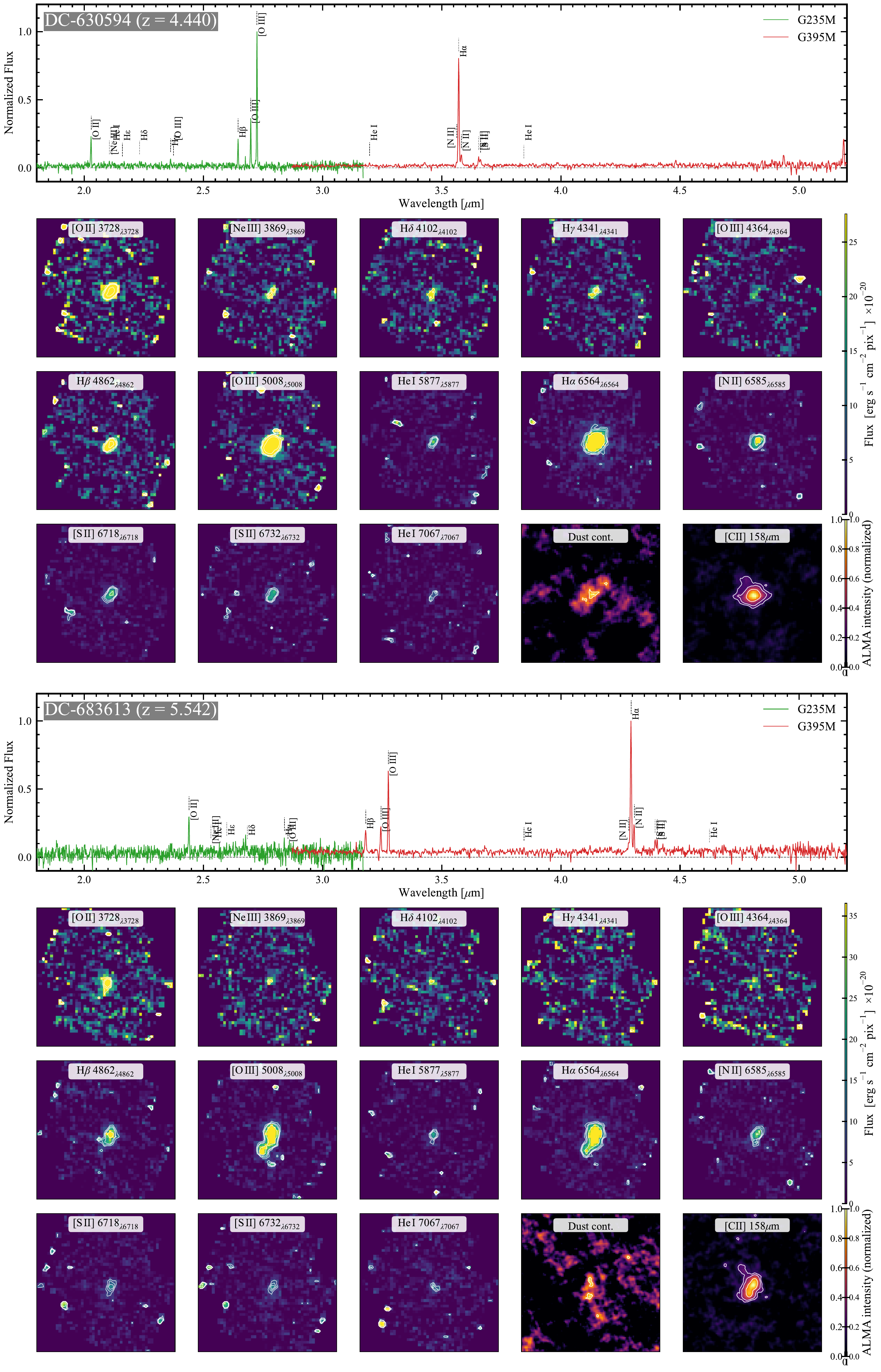}
    \caption{(continued)}
\end{figure*}

\addtocounter{figure}{-1}
\begin{figure*}
    \centering
    \includegraphics[width=0.8\textwidth]{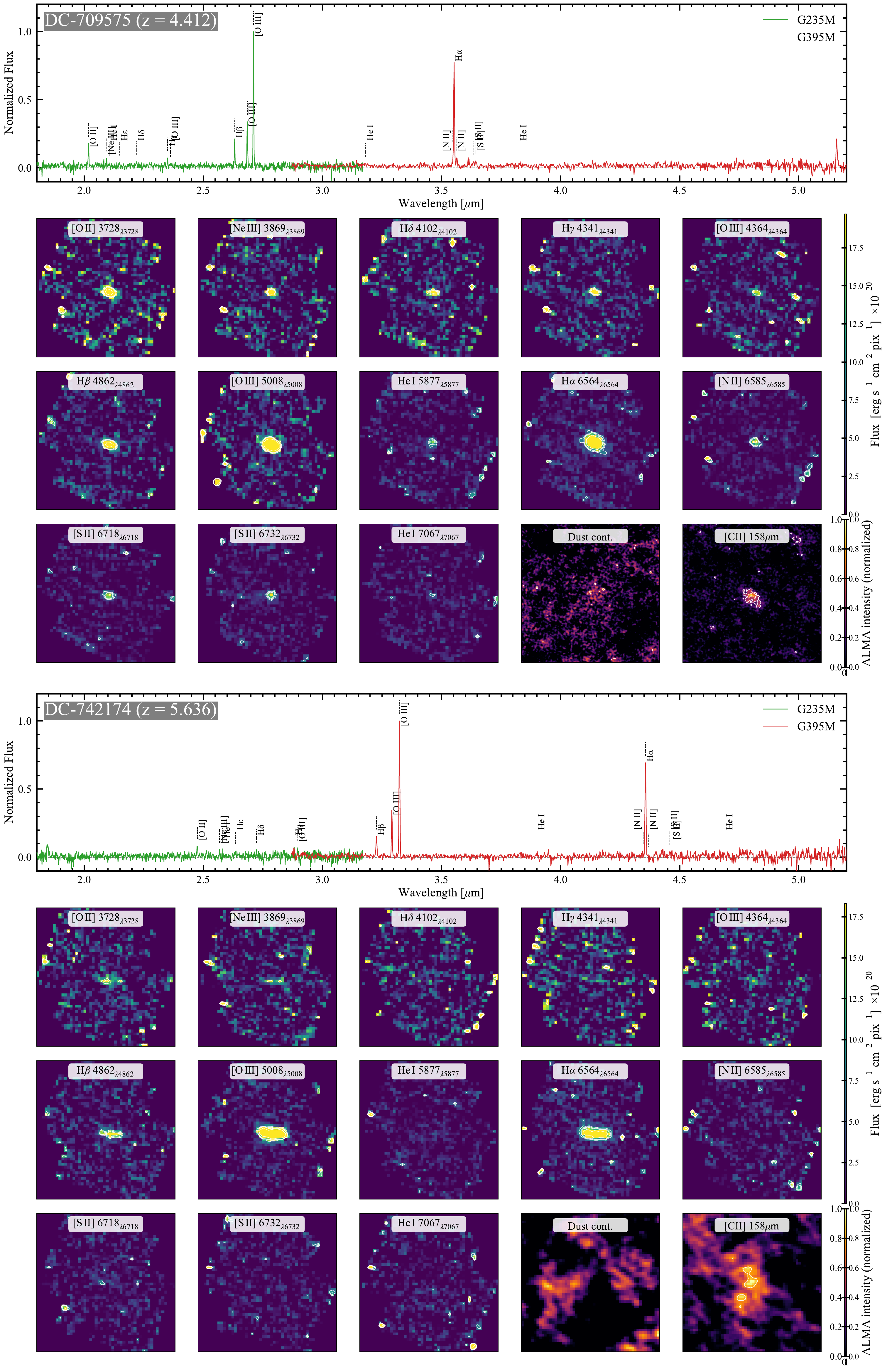}
    \caption{(continued)}
\end{figure*}

\addtocounter{figure}{-1}
\begin{figure*}
    \centering
    \includegraphics[width=0.8\textwidth]{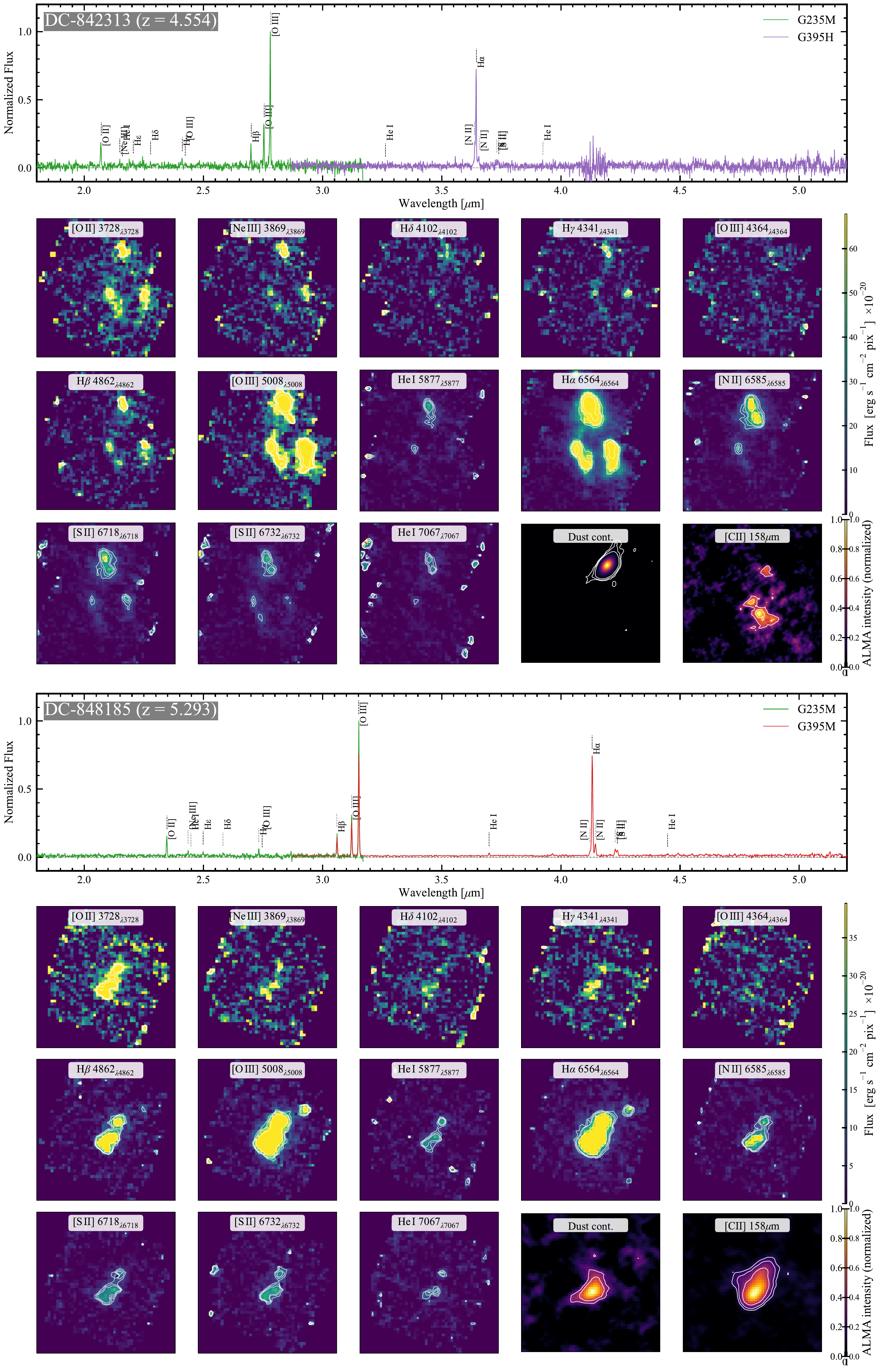}
    \caption{(continued)}
\end{figure*}

\addtocounter{figure}{-1}
\begin{figure*}
    \centering
    \includegraphics[width=0.8\textwidth]{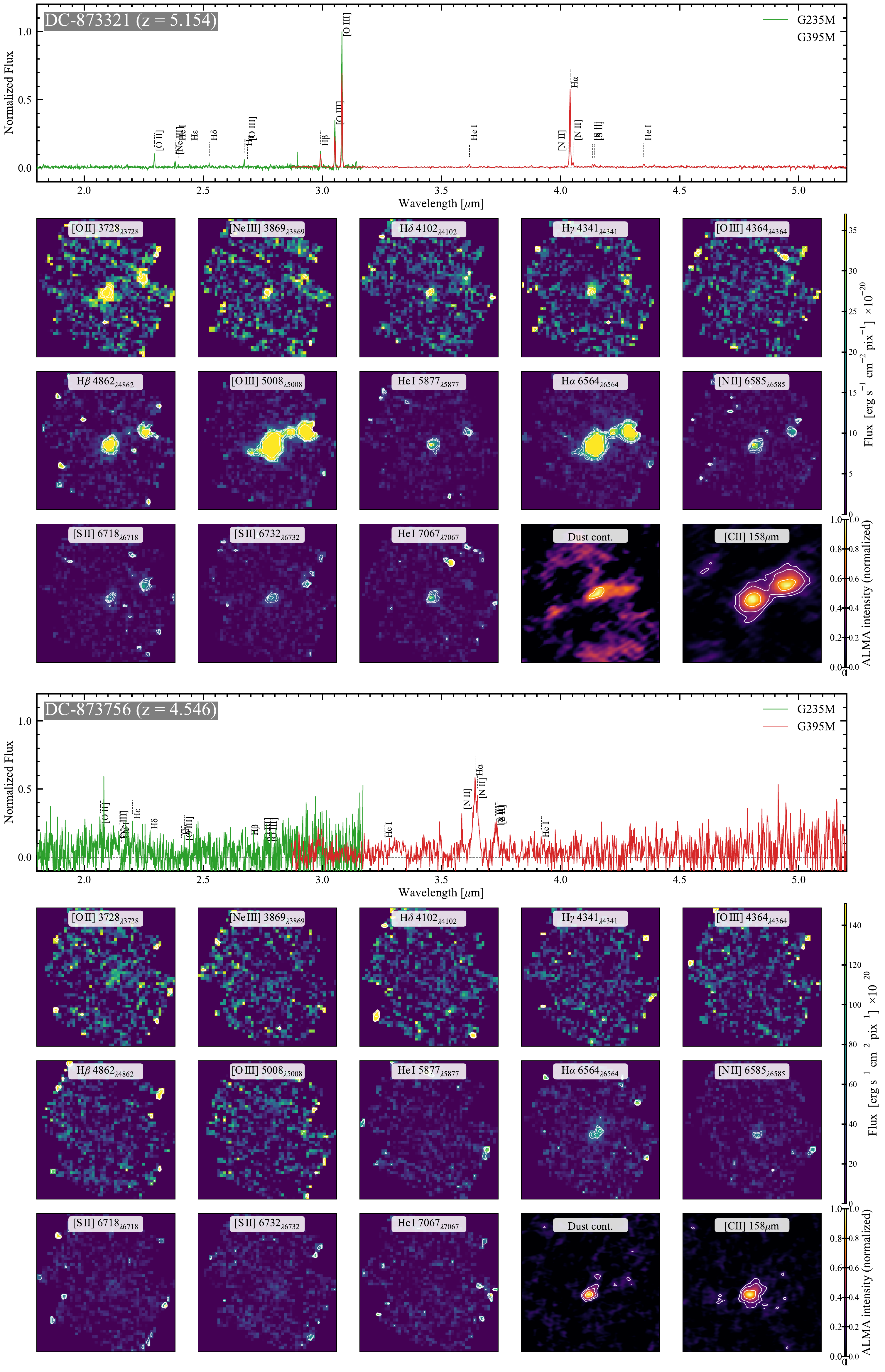}
    \caption{(continued)}
\end{figure*}

\addtocounter{figure}{-1}
\begin{figure*}
    \centering
    \includegraphics[width=0.8\textwidth]{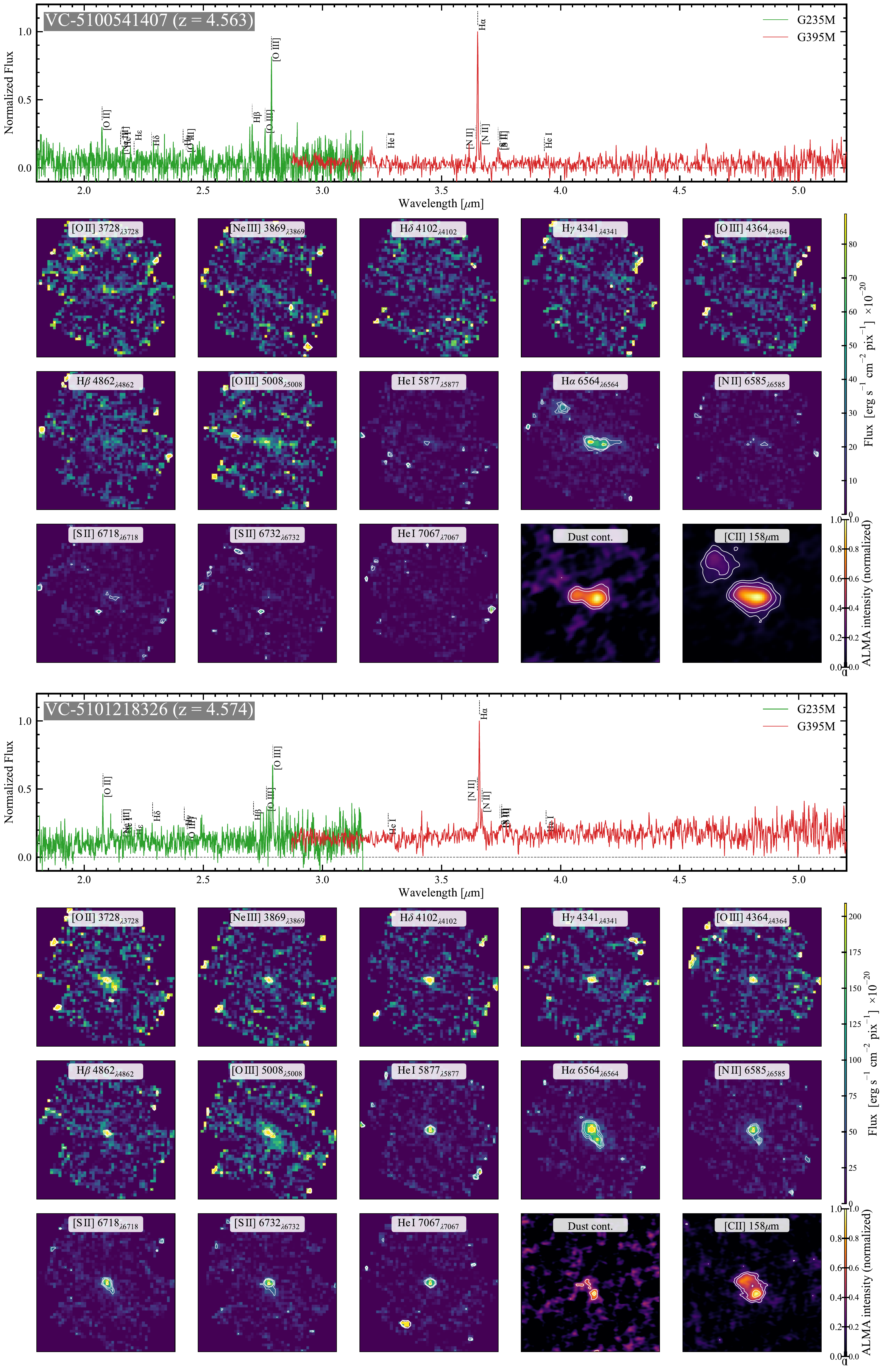}
    \caption{(continued)}
\end{figure*}

\addtocounter{figure}{-1}
\begin{figure*}
    \centering
    \includegraphics[width=0.8\textwidth]{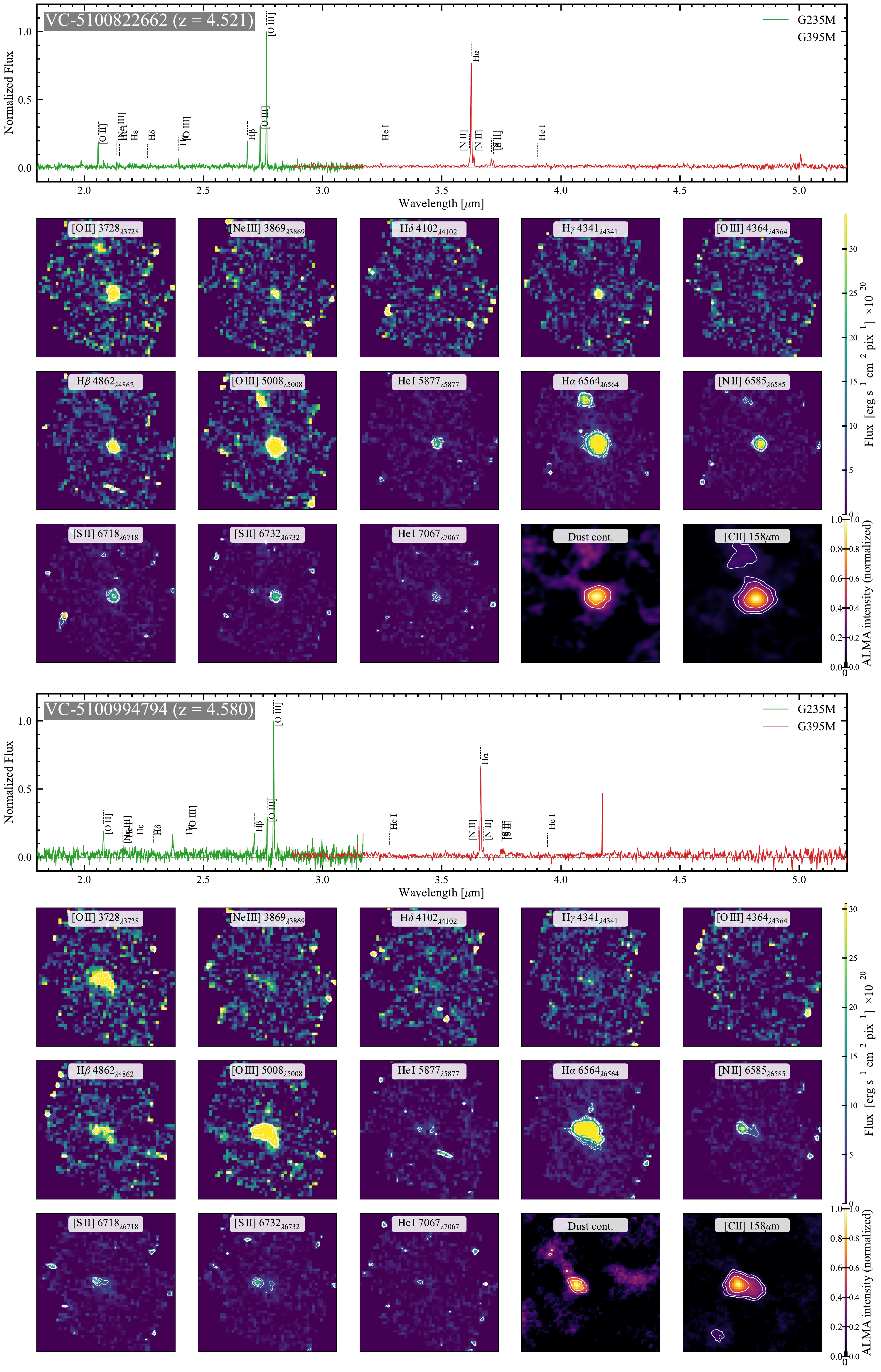}
    \caption{(continued)}
\end{figure*}

\addtocounter{figure}{-1}
\begin{figure*}
    \centering
    \includegraphics[width=0.8\textwidth]{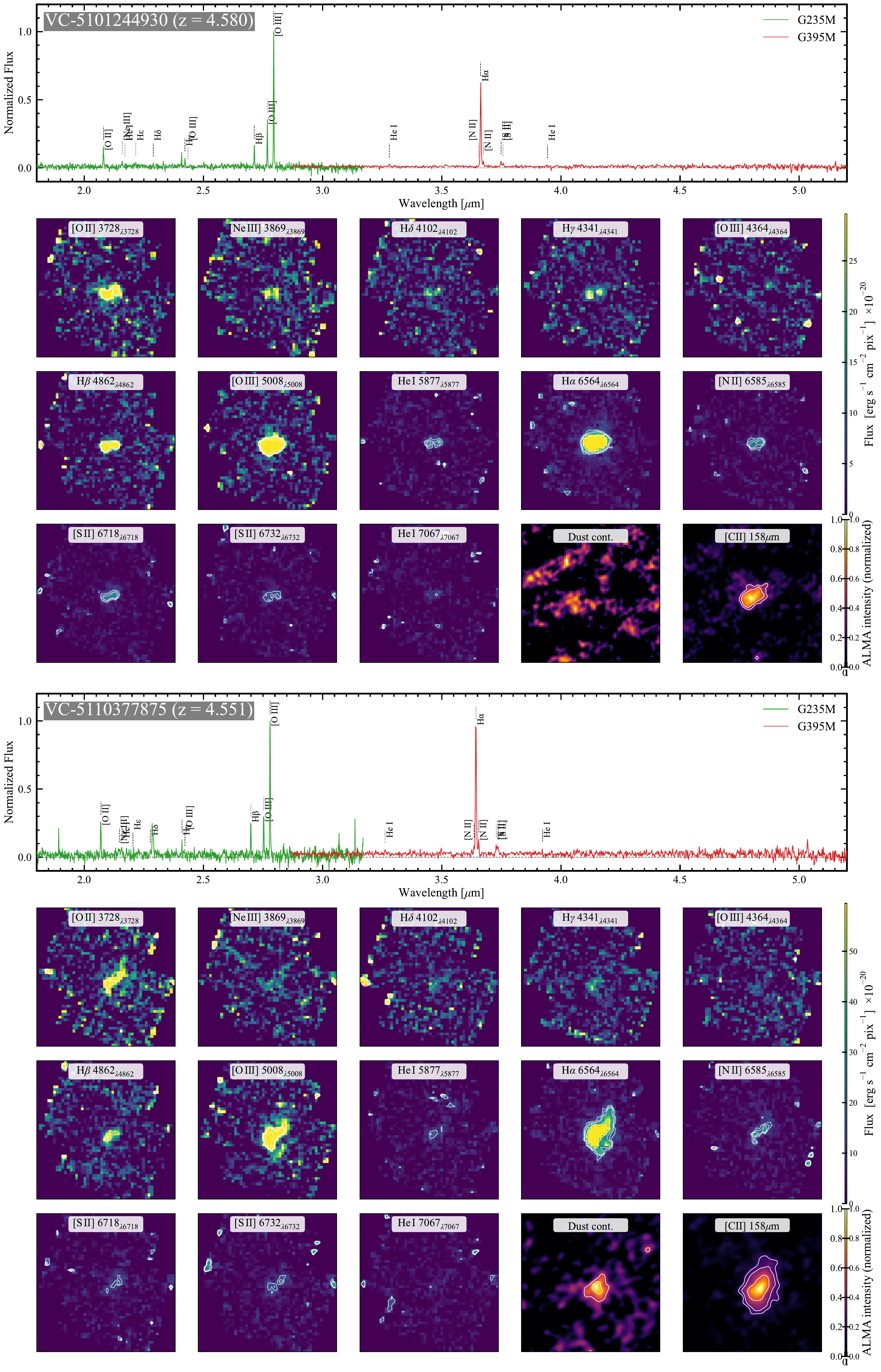}
    \caption{(continued)}
\end{figure*}

\begin{figure*}
\centering
\includegraphics[width=\textwidth]{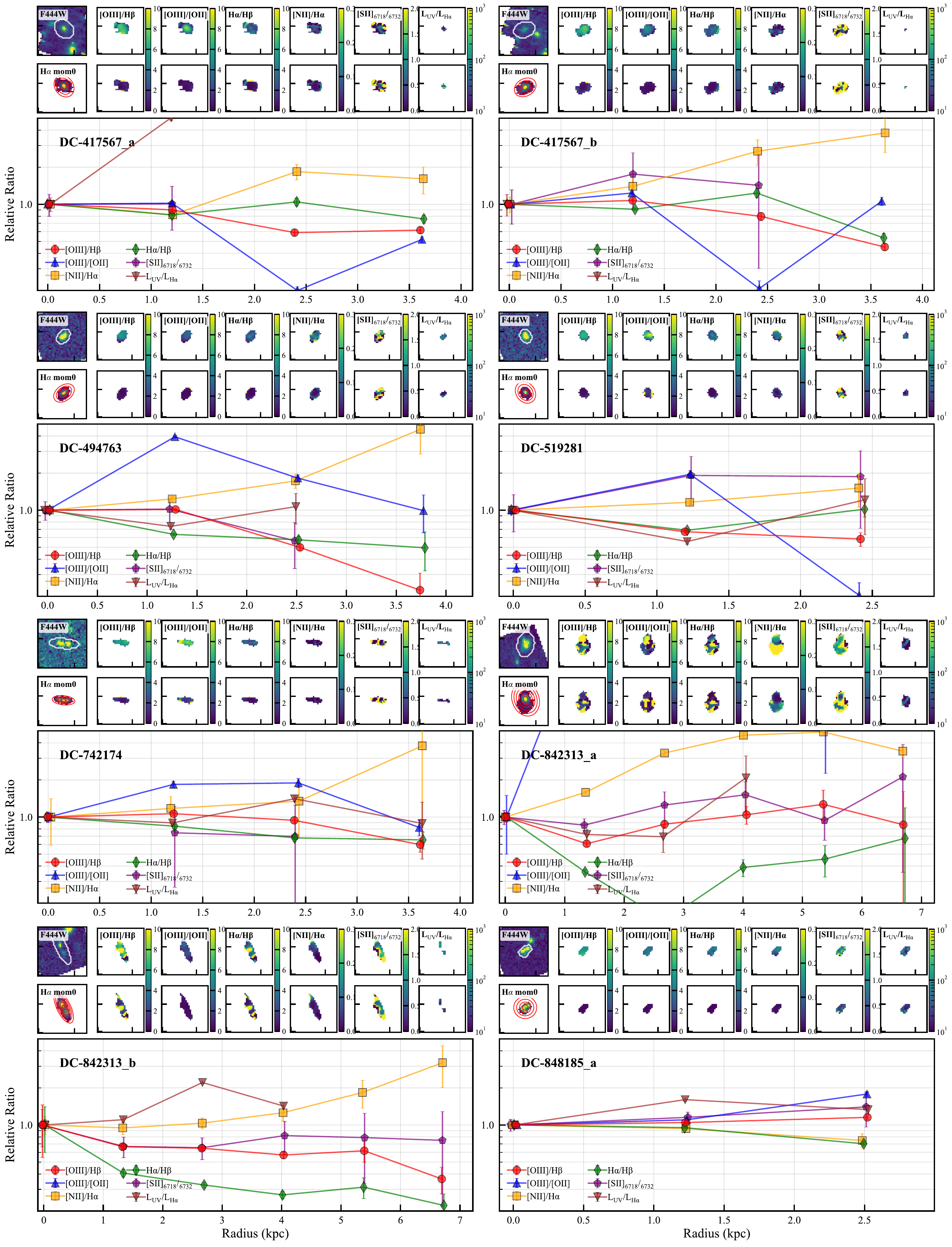}
\caption{
Same as Figure~\ref{fig:radial_ratio}, but for an additional subset of our sample. 
}
\label{fig:radial_ratios_appendix}
\end{figure*}

\addtocounter{figure}{-1}
\begin{figure*}
\centering
\includegraphics[width=\textwidth]{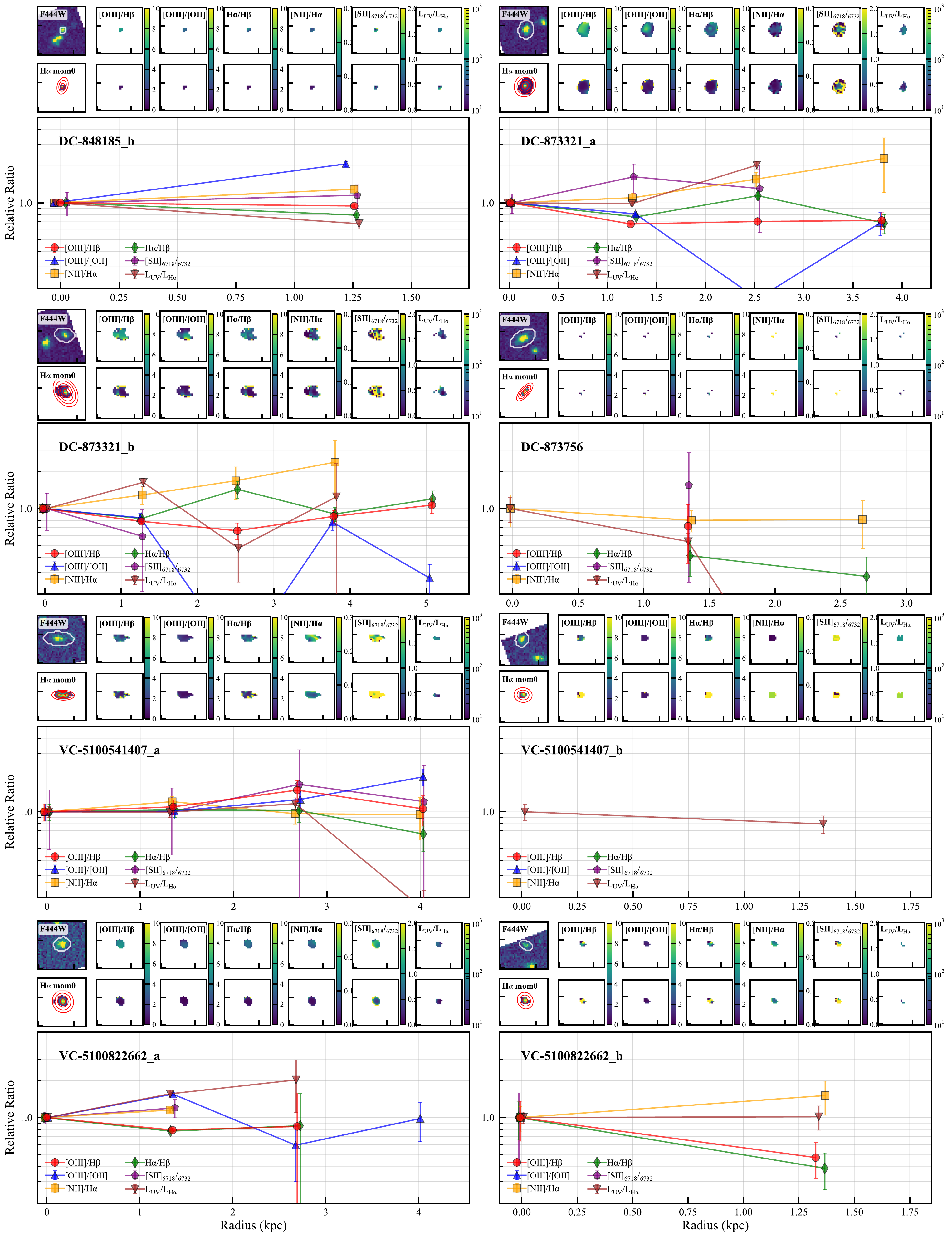}
\caption{
(Continued)
}
\end{figure*}

\addtocounter{figure}{-1}
\begin{figure*}
\centering
\includegraphics[width=\textwidth]{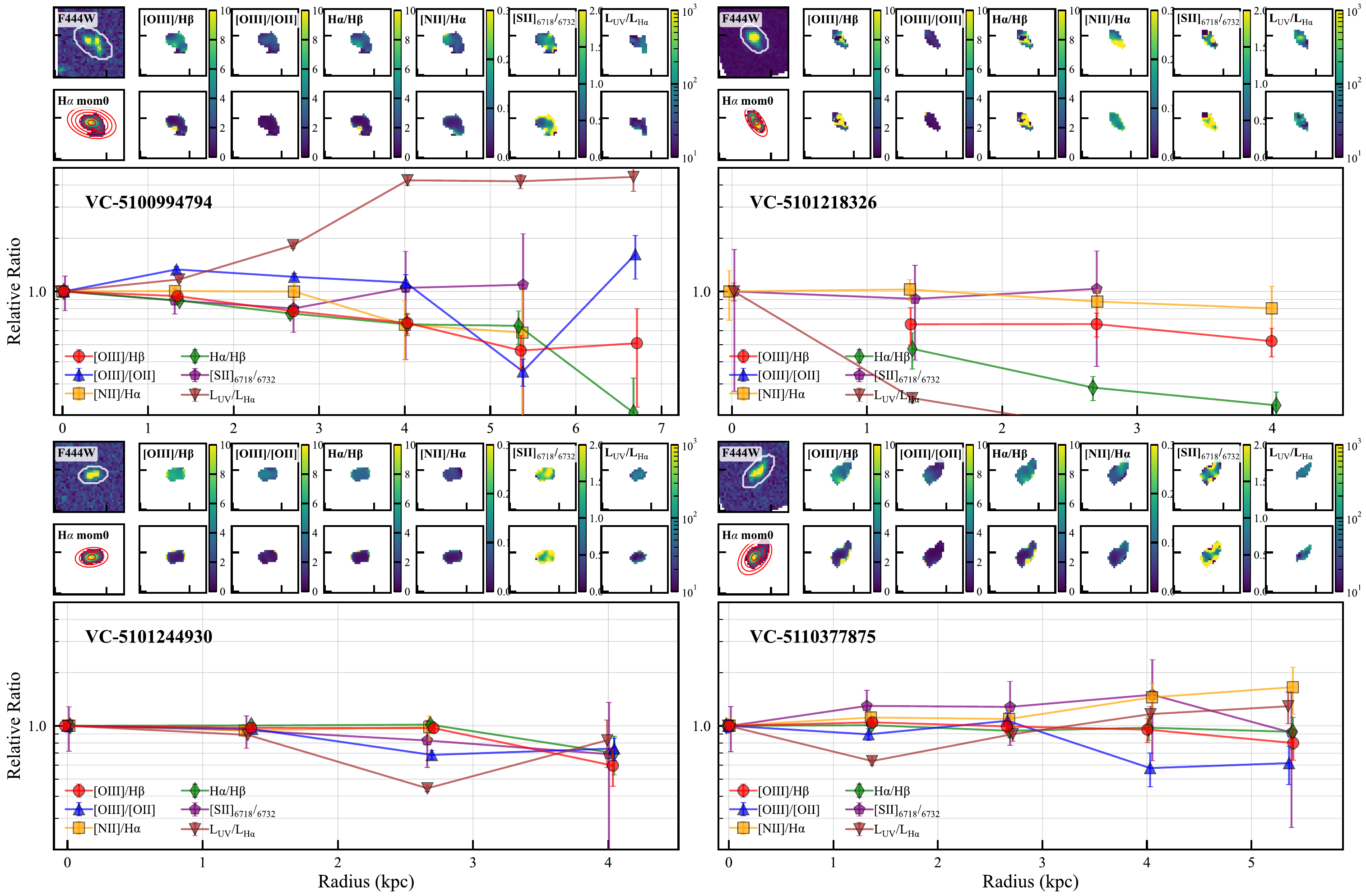}
\caption{
(Continued)
}
\end{figure*}

\bibliographystyle{apj}
\bibliography{apj-jour,reference}

\end{document}